\documentstyle[11pt,aaspp4]{article}
\begin{document}

\title{DIRECT Distances to Nearby Galaxies Using Detached Eclipsing
Binaries and Cepheids. III. Variables in the Field M31C\footnote{Based
on the observations collected at the Michigan-Dartmouth-MIT (MDM)
1.3~m telescope and at the F.~L.~Whipple Observatory (FLWO)
1.2~m telescope}}

\author{K. Z. Stanek\altaffilmark{2}} 
\affil{Harvard-Smithsonian Center for Astrophysics, 60 Garden St., MS20, 
Cambridge, MA~02138} 
\affil{\tt e-mail: kstanek@cfa.harvard.edu} 
\altaffiltext{2}{On leave from N.~Copernicus Astronomical Center, 
Bartycka 18, Warszawa PL-00-716, Poland} 
\author{J. Kaluzny}
\affil{Warsaw University Observatory, Al. Ujazdowskie 4,
PL-00-478 Warszawa, Poland} 
\affil{\tt e-mail: jka@sirius.astrouw.edu.pl} 
\author{M. Krockenberger, D. D. Sasselov} 
\affil{Harvard-Smithsonian Center for Astrophysics, 60 Garden St., MS16,
Cambridge, MA~02138} 
\affil{\tt e-mail: krocken@cfa.harvard.edu, sasselov@cfa.harvard.edu} 
\author{J. L. Tonry} 
\affil{University of Hawaii, Institute for Astronomy, 
2680 Woodlawn Dr., Honolulu, HI~96822}
\affil{\tt e-mail: jt@avidya.ifa.hawaii.edu} 
\author{M. Mateo}
\affil{Department of Astronomy, University of Michigan, 
821 Dennison Bldg., Ann Arbor, MI~48109--1090} 
\affil{\tt e-mail: mateo@astro.lsa.umich.edu}

\begin{abstract}

We undertook a long term project, DIRECT, to obtain the direct
distances to two important galaxies in the cosmological distance
ladder -- M31 and M33 -- using detached eclipsing binaries (DEBs) and
Cepheids. While rare and difficult to detect, DEBs provide us with the
potential to determine these distances with an accuracy better than
5\%. The extensive photometry obtained in order to detect DEBs
provides us with good light curves for the Cepheid variables. These
are essential to the parallel project to derive direct Baade-Wesselink
distances to Cepheids in M31 and M33. For both Cepheids and eclipsing
binaries, the distance estimates will be free of any intermediate
steps.
 
As a first step in the DIRECT project, between September 1996 and
October 1997 we obtained 95 full/partial nights on the F. L. Whipple
Observatory 1.2~m telescope and 36 full nights on the
Michigan-Dartmouth-MIT 1.3~m telescope to search for DEBs and new
Cepheids in the M31 and M33 galaxies.  In this paper, third in the
series, we present the catalog of variable stars, most of them newly
detected, found in the field M31C $[(\alpha,\delta)=
(11.\!\!\arcdeg10, 41.\!\!\arcdeg42), {\rm J2000.0}]$.  We have found
115 variable stars: 12 eclipsing binaries, 35 Cepheids and 68 other
periodic, possible long period or non-periodic variables. The catalog
of variables, as well as their photometry and finding charts, is
available via {\tt anonymous ftp} and the {\tt World Wide Web}.  The
complete set of the CCD frames is available upon request.

\end{abstract}

\keywords{binaries: eclipsing --- Cepheids --- distance scale 
--- galaxies: individual (M31) --- stars: variables: other}

\section{Introduction}

The two nearby galaxies M31 and M33 are stepping stones to most of our
current effort to understand the evolving universe at large scales.
First, they are essential to the calibration of the extragalactic
distance scale (Jacoby et al.~1992; Tonry et al.~1997). Second, they
constrain population synthesis models for early galaxy formation and
evolution and provide the stellar luminosity calibration. There is one
simple requirement for all this---accurate distances.
 
Detached eclipsing binaries (DEBs) have the potential to establish
distances to M31 and M33 with an unprecedented accuracy of better than
5\% and possibly to better than 1\%. These distances are now known to
no better than 10-15\%, as there are discrepancies of $0.2-0.3\;{\rm
mag}$ between RR Lyrae and Cepheids distance indicators (e.g.~Huterer,
Sasselov \& Schechter 1995; Holland 1998; Stanek \& Garnavich 1998).
Detached eclipsing binaries (for reviews see Andersen 1991;
Paczy\'nski 1997) offer a single step distance determination to nearby
galaxies and may therefore provide an accurate zero point
calibration---a major step towards very accurate determination of the
Hubble constant, presently an important but daunting problem for
astrophysicists.
 
The detached eclipsing binaries have yet to be used (Huterer et
al.~1995; Hilditch 1996) as distance indicators to M31 and M33.
According to Hilditch (1996), there were about 60 eclipsing binaries
of all kinds known in M31 (Gaposchkin 1962; Baade \& Swope 1963, 1965)
and only {\em one} in M33 (Hubble 1929).  Only now does the
availability of large-format CCD detectors and inexpensive CPUs make
it possible to organize a massive search for periodic variables, which
will produce a handful of good DEB candidates. These can then be
spectroscopically followed-up with the powerful new 6.5-10 meter
telescopes.

The study of Cepheids in M31 and M33 has a venerable history (Hubble
1926, 1929; Gaposchkin 1962; Baade \& Swope 1963, 1965). In the 1980s,
Freedman \& Madore (1990) and Freedman, Wilson, \& Madore (1991)
studied small samples of the earlier discovered Cepheids, to build
period-luminosity (P-L) relations in M31 and M33, respectively.
However, both the sparse photometry and the small samples do not
provide a good basis for obtaining direct Baade-Wesselink distances
(see, e.g., Krockenberger, Sasselov \& Noyes 1997) to Cepheids---the
need for new digital photometry has been long overdue. Recently,
Magnier et al.~(1997) surveyed large portions of M31, which have
previously been ignored, and found some 130 new Cepheid variable
candidates.  Their light curves are, however, rather sparsely sampled
and in the $V$-band only.

In Kaluzny et al.~(1998, hereafter: Paper I) and Stanek et al.~(1998,
hereafter: Paper II), the first two papers of the series, we presented
the catalogs of variable stars found in two fields in M31, called M31B
and M31A. Here we present the catalog of variables from the next field
M31C. In Sec.2 we discuss the selection of the fields in M31 and the
observations. In Sec.3 we describe the data reduction and
calibration. In Sec.4 we discuss briefly the automatic selection we
used for finding the variable stars. In Sec.5 we discuss the
classification of the variables.  In Sec.6 we present the catalog of
variable stars, followed by brief discussion of the results in Sec.7.

\section{Fields selection and observations}

M31 was primarily observed with the 1.3~m McGraw-Hill Telescope at the
Michigan-Dartmouth-MIT (MDM) Observatory. We used the
front-illuminated, Loral $2048^2$ CCD ``Wilbur'' (Metzger, Tonry \&
Luppino 1993), which at the $f/7.5$ station of the 1.3~m telescope has
a pixel scale of $0.32\; arcsec\; pixel^{-1}$ and field of view of
roughly $11\;arcmin$. We used Kitt Peak Johnson-Cousins $BVI$ filters.
Data for M31 were also obtained, mostly in 1997, with the 1.2~m
telescope at the F. L. Whipple Observatory (FLWO), where we used
``AndyCam'' (Szentgyorgyi et al.~1998), with a thinned, back-side
illuminated, AR coated Loral $2048^2$ pixel CCD.  The pixel scale
happens to be essentially the same as at the MDM 1.3~m telescope. We
used standard Johnson-Cousins $BVI$ filters.

Fields in M31 were selected using the MIT photometric survey of M31 by
Magnier et al.~(1992) and Haiman et al.~(1994) (see Paper I, Fig.1).
We selected six $11'\times11'$ fields, M31A--F, four of them (A--D)
concentrated on the rich spiral arm in the northeast part of M31, one
(E) coinciding with the region of M31 searched for microlensing by
Crotts \& Tomaney (1996), and one (F) containing the giant star
formation region known as NGC206 (observed by Baade \& Swope
1963). Fields A--C were observed during September and October 1996
five to eight times per night in the $V$ band, resulting in total of
110--160 $V$ exposures per field. Fields D--F were observed once a
night in the $V$-band. Some exposures in $B$ and $I$ were also
taken. M31 was also observed, in 1996 and 1997, at the FLWO 1.2~m
telescope, whose main target was M33.

In this paper we present the results for the M31C field.  We obtained
for this field useful data during 29 nights at the MDM, collecting a
total of $141\times 900\;sec$ exposures in $V$ and $30\times 600\;sec$
exposures in $I$. We also obtained for this field useful data during
24 nights at the FLWO, in 1996 and 1997, collecting a total of
$20\times 900\;sec$ exposures in $V$, $25\times 600\;sec$ exposures in
$I$ and $10\times 1200\;sec$ exposures of $B$.\footnote{The complete
list of exposures for this field and related data files are available
through {\tt anonymous ftp} on {\tt cfa-ftp.harvard.edu}, in {\tt
pub/kstanek/DIRECT} directory. Please retrieve the {\tt README} file
for instructions.  Additional information on the DIRECT project is
available through the {\tt WWW} at {\tt
http://cfa-www.harvard.edu/\~\/kstanek/DIRECT/}.}

\section{Data reduction, calibration and astrometry}

The details of the reduction procedure were given in Paper I.
Preliminary processing of the CCD frames was done with the standard
routines in the IRAF-CCDPROC package.\footnote{IRAF is distributed by
the National Optical Astronomy Observatories, which are operated by
the Associations of Universities for Research in Astronomy, Inc.,
under cooperative agreement with the NSF} Stellar profile photometry
was extracted using the {\it Daophot/Allstar} package (Stetson 1987,
1992).  We selected a ``template'' frame for each filter using a
single frame of particularly good quality.  These template images were
reduced in a standard way (Paper I).  Other images were reduced using
{\it Allstar} in the fixed-position-mode using as an input the
transformed object list from the template frames.  For each frame the
list of instrumental photometry derived for a given frame was
transformed to the common instrumental system of the appropriate
``template'' image.  Photometry obtained for the $B,V$ and $I$ filters
was combined into separate data bases. M31C images obtained at the
FLWO were reduced using MDM ``templates''.  In case of $B$-band images
obtained at FLWO we used the $V$-band MDM template to fix the
positions of the stars.

The photometric $VI$ calibration of the MDM data was discussed in
Paper I.  In addition, for the field M31C on the night of 1997 October
9/10 we have obtained independent $BVI$ calibration with the FLWO
1.2~m telescope. There was an offset of $0.012\;{\rm mag}$ in $V$ and
$0.024\;{\rm mag}$ in $V-I$ between the FLWO and the MDM calibration,
i.e. well within our estimate of the total $0.05\;mag$ systematic
error discussed in Paper I.
We also
derived equatorial coordinates for all objects included in the data
bases for the $V$ filter. The transformation from rectangular
coordinates to equatorial coordinates was derived using $\sim200$
stars identified in the list published by Magnier et al.~(1992).

\section{Selection of variables}

\begin{figure}[t]
\plotfiddle{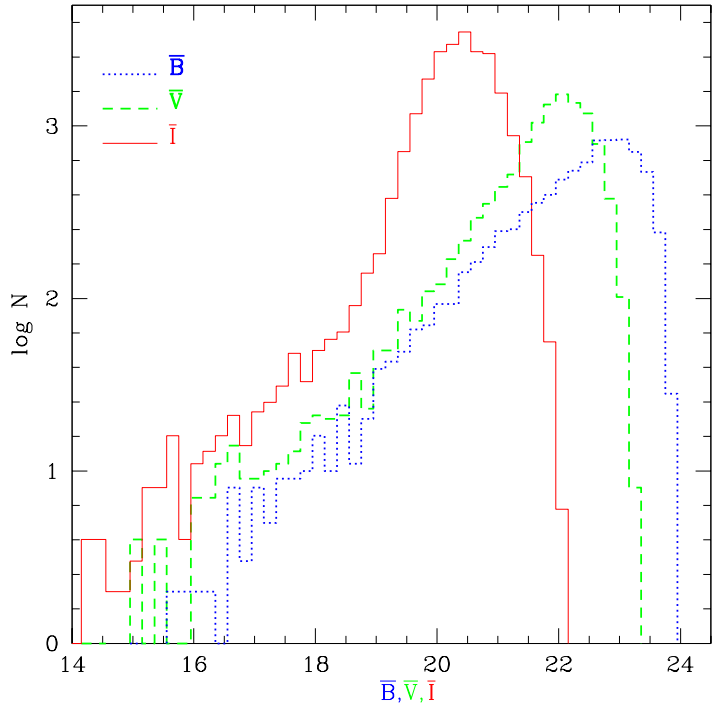}{8cm}{0}{50}{50}{-160}{-85}
\caption{Distributions in $B$ (dotted line), $V$ (dashed line) and $I$
(continuous line) of stars in the field M31C.}
\label{fig:dist}
\end{figure}

The procedure for selecting the variables was described in detail in
Paper I, so here we only give a short description, noting changes when
necessary.  The reduction procedure described in previous section
produces databases of calibrated $BVI$ magnitudes and their standard
errors. The $BV$ databases for M31C field contain 15120 stars, with up
to 161 measurements in $V$ and up to 10 measurements in $B$, and the
$I$ database contains 28441 stars with up to 55 measurements.
Figure~\ref{fig:dist} shows the distributions of stars as a function of
mean $\bar{B}$, $\bar{V}$ or $\bar{I}$ magnitude.  As can be seen from
the shape of the histograms, our completeness starts to drop rapidly
at about $\bar{B}\sim23$, $\bar{V}\sim22$ and $\bar{I}\sim20.5$. The
primary reason for this difference in the depth of the photometry
between $BV$ and $I$ is the level of the combined sky and background
light, which is about three times higher in the $I$ filter than in the
$BV$ filters.

The measurements flagged as ``bad'' and measurements with errors
exceeding the average error by more than $4\sigma$ are removed
(Paper~I).  Usually zero to 10 points are removed, leaving the
majority of stars with roughly $N_{good}\sim150-160$ $V$\/
measurements.  For further analysis we use only those stars that have
at least $N_{good}>N_{max}/2\;(=80)$ measurements. There are 11263
such stars in the $V$ database of the M31C field.

\begin{figure}[t]
\plotfiddle{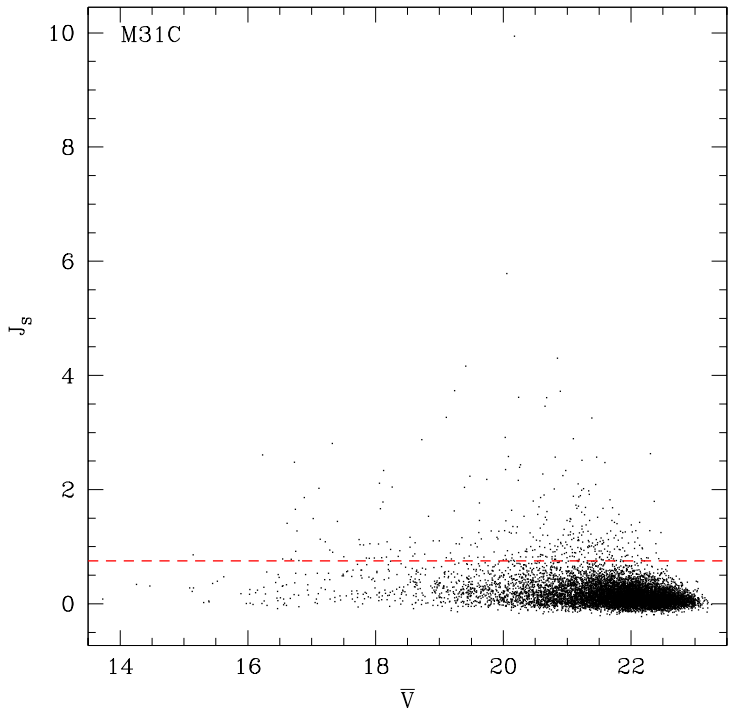}{8cm}{0}{50}{50}{-160}{-85}
\caption{Variability index $J_S$ vs. mean $\bar{V}$ magnitude for 11262
stars in the field M31C with $N_{good}>80$.  Dashed line at $J_S=0.75$
defines the cutoff applied for variability.}
\label{fig:stetj}
\end{figure}

Our next goal is to select objectively a sample of variable stars from
the total sample defined above.  There are many ways to proceed, and
we largely follow the approach of Stetson (1996).  The procedure is
described in more detail in Paper~I. In short, for each star we
compute the Stetson's variability index $J_S$ (Paper I, Eq.7), and
stars with values exceeding some minimum value $J_{S,min}$ are
considered candidate variables.  The definition of Stetson's
variability index includes the standard errors of individual
observations.  If, for some reason, these errors were over- or
underestimated, we would either miss real variables, or select
spurious variables as real ones. Using the procedure described in
Paper I, we scale the {\em Daophot} errors to better represent the
``true'' photometric errors.  We then select the candidate variable
stars by computing the value of $J_S$ for the stars in our $V$
database.  We used a cutoff of $J_{S,min}=0.75$ and additional cuts
described in Paper I to select 313 candidate variable stars (about 3\%
of the total number of 11263).  In Figure~\ref{fig:stetj} we plot the
variability index $J_S$ vs. apparent visual magnitude $\bar{V}$ for
11262 stars with $N_{good}>80$.

\section{Period determination, classification of variables}

We based our candidate variables selection on the $V$ band data
collected at the MDM and the FLWO telescopes.  We also have the
$BI$-bands data for the field, up to 55 $I$-band epochs and up to 10
$B$-band epochs, although for a variety of reasons some of the
candidate variable stars do not have an $B$ or $I$-band
counterpart. We will therefore not use the $BI$ data for the period
determination and broad classification of the variables. We will
however use the $BI$ data for the ``final'' classification of some
variables.

Next we searched for the periodicities for all 313 candidate
variables, using a variant of the Lafler-Kinman (1965) technique
proposed by Stetson (1996). Starting with the minimum period of
$0.25\;days$, successive trial periods are chosen so
\begin{equation}
P_{j+1}^{-1}=P_{j}^{-1}-\frac{0.02}{\Delta t},
\end{equation}
where $\Delta t=t_{N}-t_{1}=398\;days$ is the time span of the series.
The maximum period considered is $150\;days$.  For each candidate
variable 10 best trial periods are selected (Paper I) and then used in
our classification scheme.

The variables we are most interested in are Cepheids and eclipsing
binaries (EBs). We therefore searched our sample of variable stars for
these two classes of variables. As mentioned before, for the broad
classification of variables we restricted ourselves to the $V$ band
data.  We will, however, present and use the $BI$-bands data, when
available, when discussing some of the individual variable stars.

For EBs we used search strategy described in Paper II.  Within our
assumption the light curve of an EB is determined by nine parameters:
the period, the zero point of the phase, the eccentricity, the
longitude of periastron, the radii of the two stars relative to the
binary separation, the inclination angle, the fraction of light coming
from the bigger star and the uneclipsed magnitude. A total of 17
variables passed all of the criteria. We then went back to the CCD
frames and tried to see by eye if the inferred variability is indeed
there, especially in cases when the light curve is very noisy/chaotic.
We decided to remove five dubious eclipsing binaries.  The remaining
12 EBs with their parameters and light curves are presented in the
Section~6.1.

In the search for Cepheids we followed the approach by Stetson (1996)
of fitting template light curves to the data. We used the
parameterization of Cepheid light curves in the $V$-band as given by
Stetson (1996). There was a total of 100 variables passing all of the
criteria (Paper I and II), but after investigating the CCD frames we
removed 28 dubious ``Cepheids'', which leaves us with 62 probable
Cepheids. Their parameters and light curves are presented in the
Sections~6.2,~6.3.

After the preliminary selection of 17 eclipsing binaries and 100
possible Cepheids, we were left with 197 ``other'' variable stars.
After raising the threshold of the variability index to
$J_{S,min}=1.2$ (Paper I) we are left with 61 variables. After
investigating the CCD frames we removed 30 dubious variables from the
sample, which leaves 31 variables which we classify as
miscellaneous. Their parameters and light curves are presented in the
Section~6.4.

\section{Catalog of variables}

In this section we present light curves and some discussion of the 115
variable stars discovered by our survey in the field M31C.
\footnote{Complete $V$ and (when available) $BI$ photometry and
$128\times128\;pixel$ ($\sim 40''\times40''$) $V$ finding charts for
all variables are available from the authors via the {\tt anonymous
ftp} from the Harvard-Smithsonian Center for Astrophysics and can be
also accessed through the {\tt World Wide Web}.}  The variable stars
are named according to the following convention: letter V for
``variable'', the number of the star in the $V$ database, then the
letter ``D'' for our project, DIRECT, followed by the name of the
field, in this case (M)31C, e.g. V9037 D31C.  Tables~\ref{table:ecl},
\ref{table:ceph}, \ref{table:per} and \ref{table:misc} list the
variable stars sorted broadly by four categories: eclipsing binaries,
Cepheids, other periodic variables and ``miscellaneous'' variables, in
our case meaning ``variables with no clear periodicity''. Some of the
variables which were found independently by survey of Magnier et
al.~(1997) are denoted in the ``Comments'' by ``Ma97 ID'', where the
``ID'' is the identification number assigned by Magnier at
al.~(1997). We also cross-identify several variables found by us in
Paper I.

\subsection{Eclipsing binaries}

In Table~\ref{table:ecl} we present the parameters of the 12 eclipsing
binaries in the M31C field.  The lightcurves of these variables are
shown in Figure~\ref{fig:ecl}, along with the simple eclipsing binary
models discussed in the Paper I.  The variables are sorted in the
Table~\ref{table:ecl} by the increasing value of the period $P$. For
each eclipsing binary we present its name, J2000.0 coordinates (in
degrees), period $P$, magnitudes $V_{max}, I_{max}$ and $B_{max}$ of
the system outside of the eclipse, and the radii of the binary
components $R_1,\;R_2$ in the units of the orbital separation.  We
also give the inclination angle of the binary orbit to the line of
sight $i$ and the eccentricity of the orbit $e$. The reader should
bear in mind that the values of $V_{max},\;I_{max},\;B_{max},\;
R_1,\;R_2,\;i$ and $e$ are derived with a straightforward model of the
eclipsing system, so they should be treated only as reasonable
estimates of the ``true'' value.

\begin{small}
\tablenum{1} 
\begin{planotable}{lrrrrrrrrcrl}
\tablewidth{45pc}
\tablecaption{\sc DIRECT Eclipsing Binaries in M31C}
\tablehead{ \colhead{Name} & \colhead{$\alpha_{J2000.0}$} & \colhead{$\delta_{J2000.0}$} 
& \colhead{$P$} & \colhead{} & \colhead{} & \colhead{} &
\colhead{} & \colhead{} & \colhead{$i$} & \colhead{} & \colhead{} \\ 
\colhead{(D31C)} & \colhead{$(\deg)$} & \colhead{$(\deg)$}
& \colhead{$(days)$} & \colhead{$V_{max}$}  & \colhead{$I_{max}$} 
& \colhead{$B_{max}$} & \colhead{$R_1$} & \colhead{$R_2$} & \colhead{(deg)}  
& \colhead{$e$} & \colhead{Comments} } 
\startdata
V12262\ldots   & 11.1503 & 41.4888 &  2.0489 & 20.02 & 20.19 & 19.85 & 0.50 & 0.32 & 71 & 0.00 & \nl
V12594\dotfill & 11.1568 & 41.4962 &  2.3013 & 20.49 & 20.68 & 20.37 & 0.59 & 0.41 & 72 & 0.01 & V2763 D31B\nl
V10732\dotfill & 11.1246 & 41.3909 &  2.3048 & 20.65 &\nodata& 20.34 & 0.42 & 0.34 & 84 & 0.00 & DEB \nl
V14662\dotfill & 11.2087 & 41.4686 &  2.8606 & 20.28 & 19.81 & 20.19 & 0.46 & 0.35 & 83 & 0.01 & DEB \nl
V10550\dotfill & 11.1219 & 41.3837 &  3.1687 & 19.35 & 19.26 & 19.18 & 0.50 & 0.49 & 90 & 0.00 & \nl
V12650\dotfill & 11.1582 & 41.4899 &  3.5500 & 19.19 & 18.96 & 19.13 & 0.50 & 0.49 & 89 & 0.02 & \nl
V14653\dotfill & 11.2081 & 41.4807 &  3.8839 & 20.60 &\nodata& 20.36 & 0.50 & 0.49 & 90 & 0.02 & \nl
V14396\dotfill & 11.2035 & 41.3833 &  5.4260 & 21.32 &\nodata& 21.80 & 0.63 & 0.37 & 90 & 0.00 & \nl
V9037\dotfill  & 11.0969 & 41.4523 &  5.7735 & 19.22 & 19.13 & 19.24 & 0.31 & 0.26 & 76 & 0.14 & DEB \nl
V11295\dotfill & 11.1333 & 41.4223 &  7.6907 & 17.31 & 16.16 & 18.20 & 0.52 & 0.39 & 49 & 0.00 & W UMa \nl
V14439\dotfill & 11.2024 & 41.4577 &  9.1370 & 21.12 & 20.96 & 21.06 & 0.33 & 0.33 & 72 & 0.17 & DEB? \nl
V13944\dotfill & 11.1886 & 41.4667 & 11.5385 & 18.68 & 18.55 & 18.71 & 0.68 & 0.32 & 69 & 0.00 & Ma97 92
\enddata 
\tablecomments{V9037 D31C with period $P=5.7735\;days$ is a good
detached eclipsing binary (DEB) candidate, with significant
eccentricity. V12594 D31C was found in Paper I as V2763 D31B, with
$P=2.302\;days$, $V_{max}=20.51$ and $I_{max}=20.84$.}
\label{table:ecl}
\end{planotable}
\end{small}

One of the eclipsing binaries found, V9037 D31C, is a very good DEB
candidate, with deep eclipses and the ellipticity indicating that the
system is young and unevolved. However, much better light curve is
necessary to accurately establish the properties of the system.  Two
other systems, V10732 and V14662 D31C, also seem to be detached, but
they are significantly fainter than V9037 D31C and therefore less
suitable for follow-up.

Inspection of the $V,\;B-V$ color-magnitude diagram
(Figure~\ref{fig:cmd}) reveals that one of the candidate eclipsing
binaries lands close to the Cepheid portion of the CMD.  It turns out
that this variable, V14396 D31C, is only marginally better fit by a
eclipsing binary light curve than by a Cepheid light curve with
roughly half of the period, but we decided to keep it classified as an
eclipsing binary.

\begin{figure}[p]
\plotfiddle{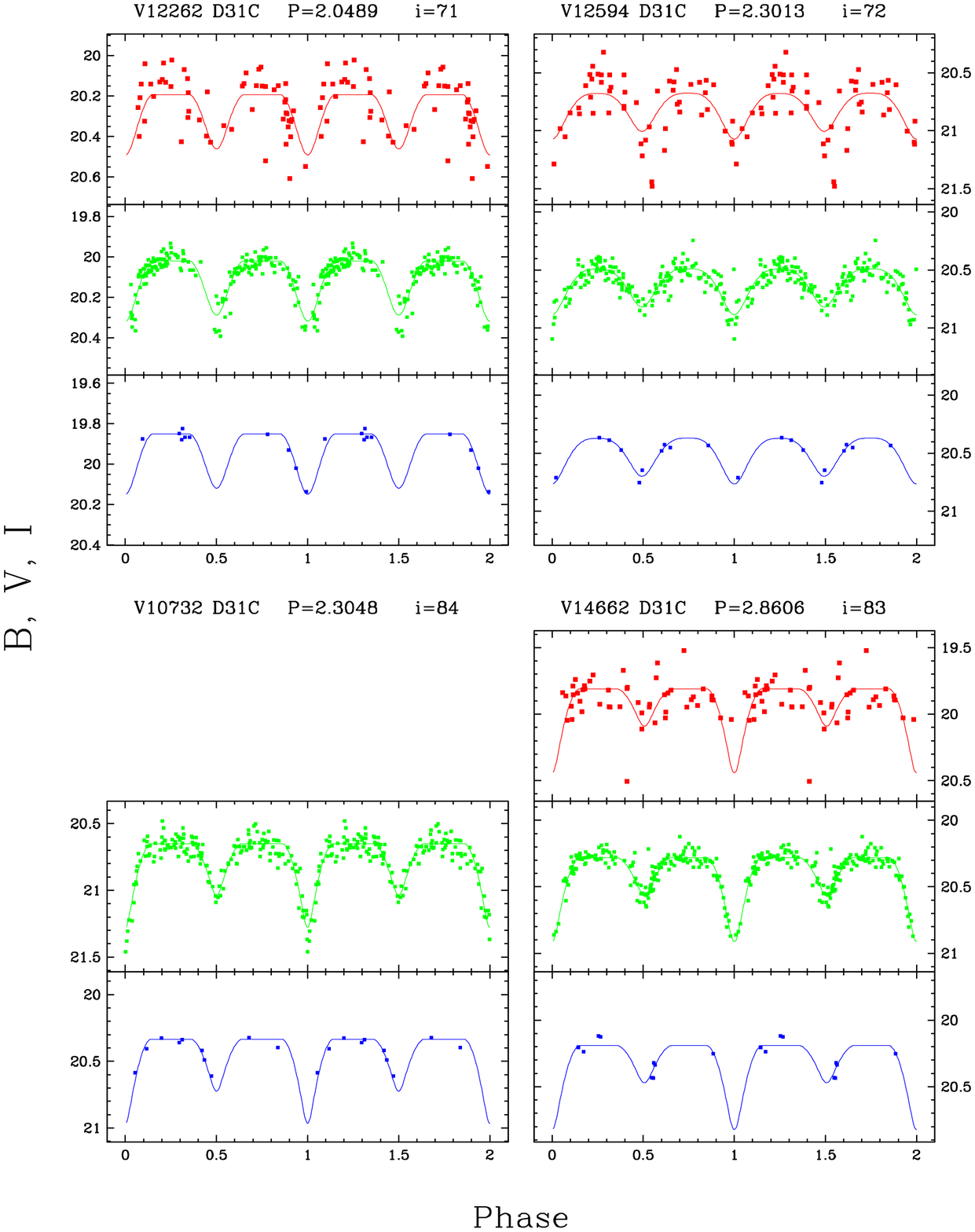}{19.5cm}{0}{83}{83}{-260}{-40}
\caption{$BVI$ lightcurves of eclipsing binaries found in the field
M31C. The thin continuous line represents the best fit model for each
star and photometric band. $B$-band lightcurve is shown in the bottom
panel and $I$-band lightcurve (when present) is shown in the top
panel.}
\label{fig:ecl}
\end{figure}
 
\addtocounter{figure}{-1}
\begin{figure}[p]
\plotfiddle{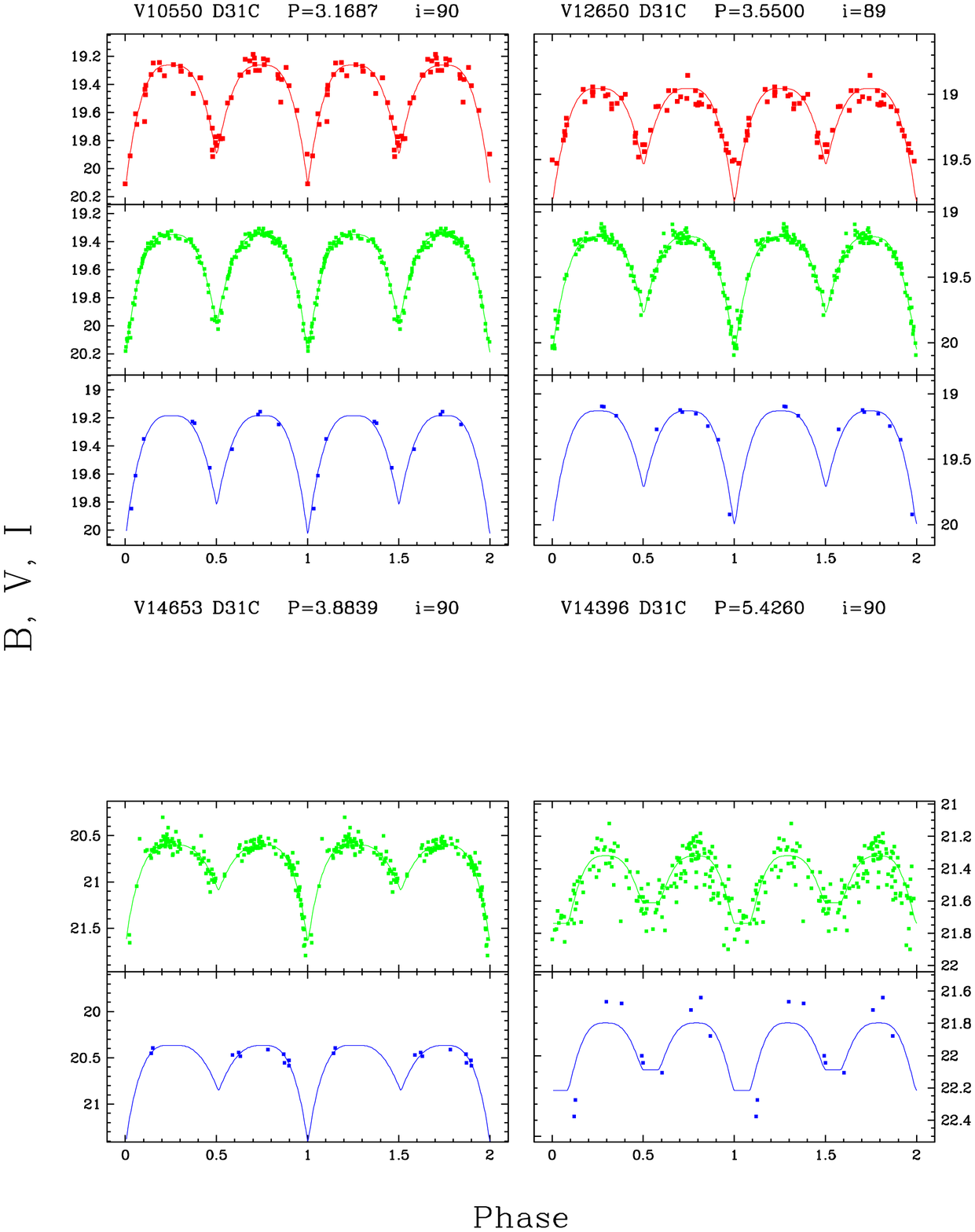}{19.5cm}{0}{83}{83}{-260}{-40}
\caption{Continued.}
\end{figure}
 
\addtocounter{figure}{-1}
\begin{figure}[p]
\plotfiddle{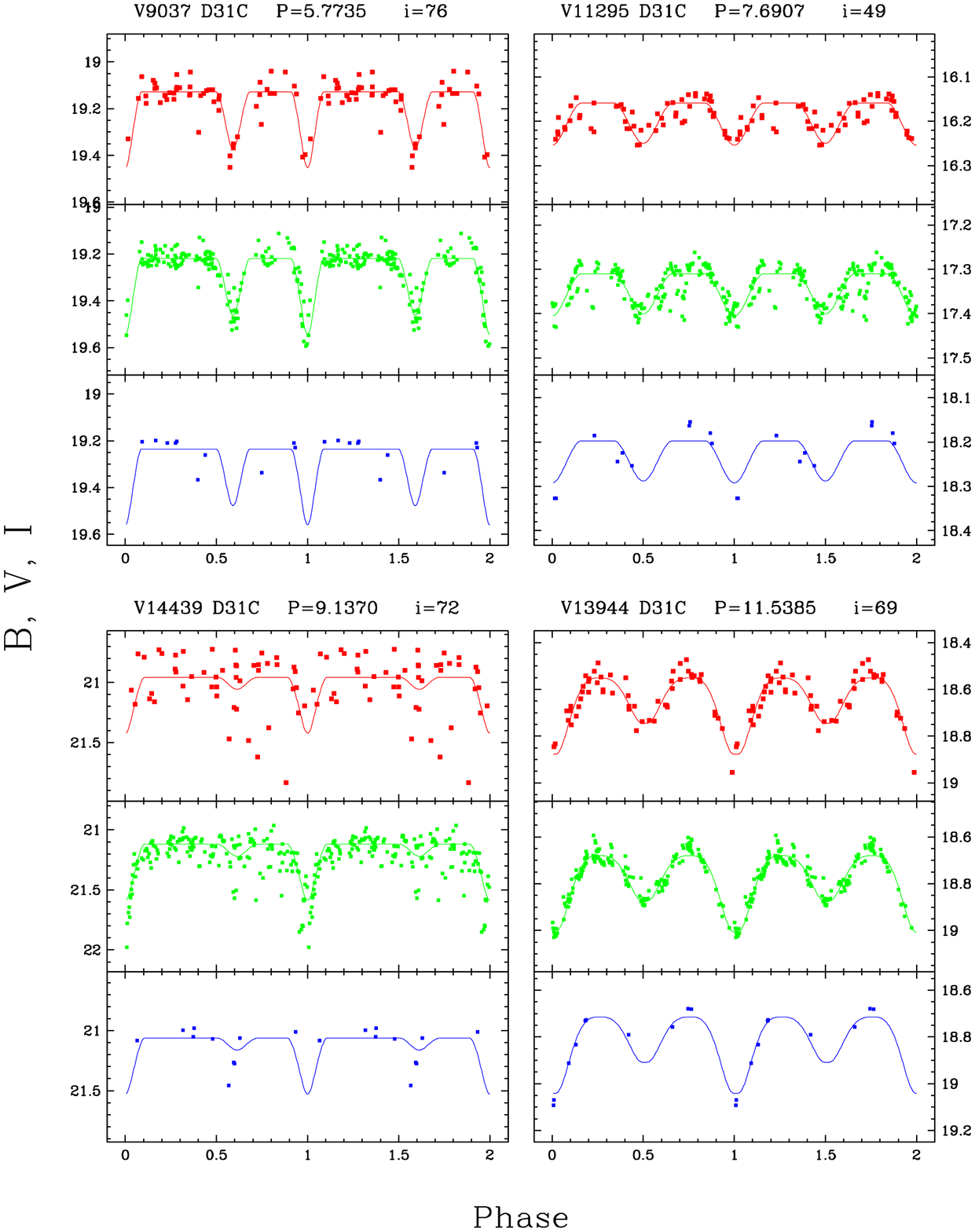}{19.5cm}{0}{83}{83}{-260}{-40}
\caption{Continued.}
\end{figure}

\begin{figure}[p]
\plotfiddle{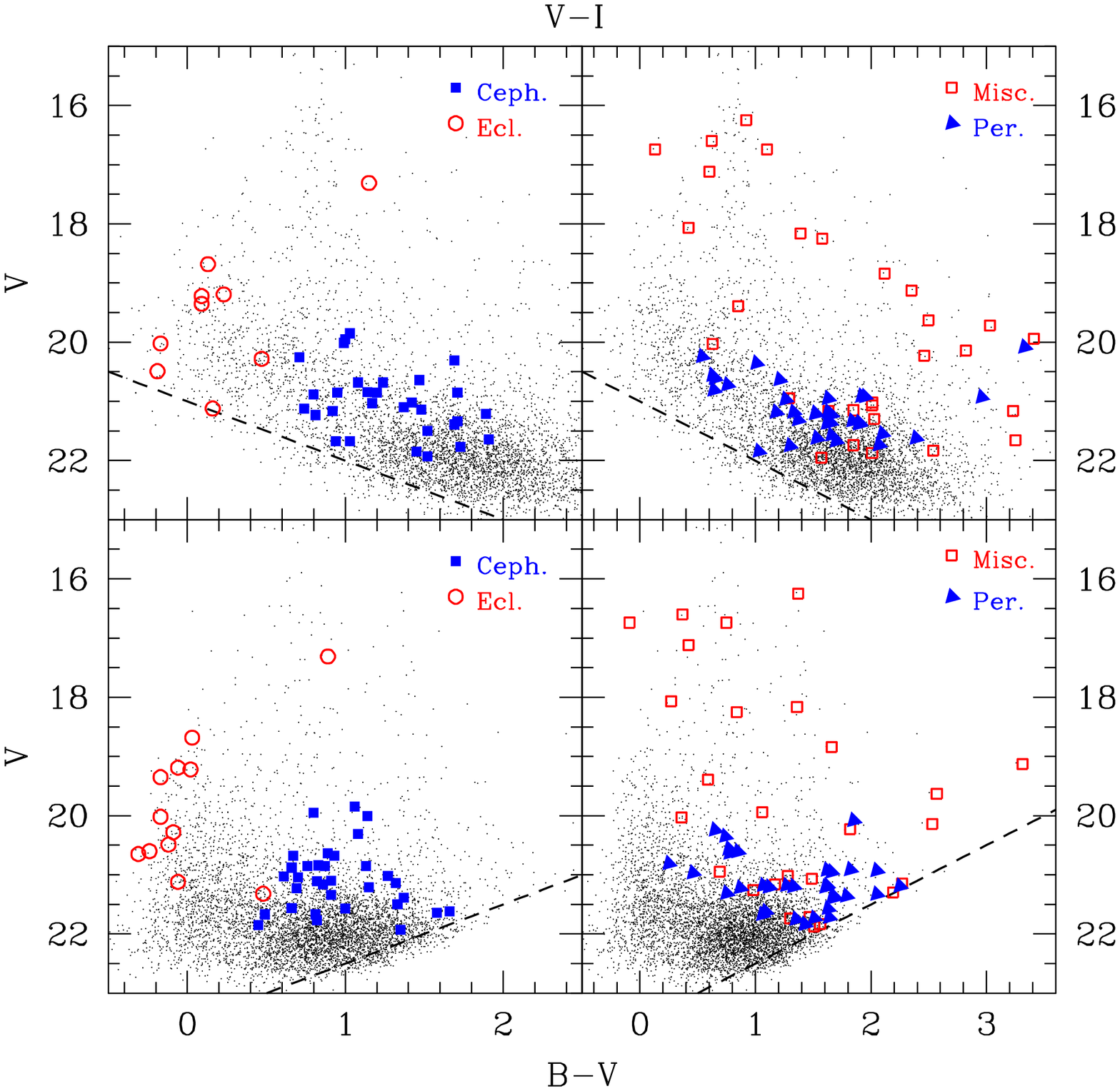}{13cm}{0}{85}{85}{-260}{-135}
\caption{$V,\;V-I$ (upper panels) and $B,\;B-V$ (lower panels)
color-magnitude diagrams for the variable stars found in the field
M31C. The eclipsing binaries and Cepheids are plotted in the left
panels and the other periodic variables and miscellaneous variables
are plotted in the right panels. The dashed lines correspond to the
$I$ detection limit of $I\sim21\;{\rm mag}$ (upper panels) and the $B$
detection limit of $B\sim23.5\;{\rm mag}$ (lower panels).
\label{fig:cmd}}
\end{figure}

\subsection{Cepheids}

In Table~\ref{table:ceph} we present the parameters of 35 Cepheids in
the M31C field, sorted by the period $P$.  For each Cepheid we present
its name, J2000.0 coordinates, period $P$, flux-weighted mean
magnitudes $\langle V\rangle$ and (when available) $\langle I\rangle$
and $\langle B\rangle$, and the $V$-band amplitude of the variation
$A$.  In Figure~\ref{fig:ceph} we show the phased $B,V,I$ lightcurves
of our Cepheids. Also shown is the best fit template lightcurve
(Stetson 1996), which was fitted to the $V$ data and then for the $I$
data only the zero-point offset was allowed. For the $B$-band data,
lacking the template lightcurve parameterization (Stetson 1996), we
used the $V$-band template, allowing for different zero-points and
amplitudes. With our limited amounts of $B$-band data this approach
produces mostly satisfactory results, but extending the
template-fitting approach of Stetson (1996) to the $B$-band (and
possibly other popular bands) would be most useful.

\begin{figure}[p]
\plotfiddle{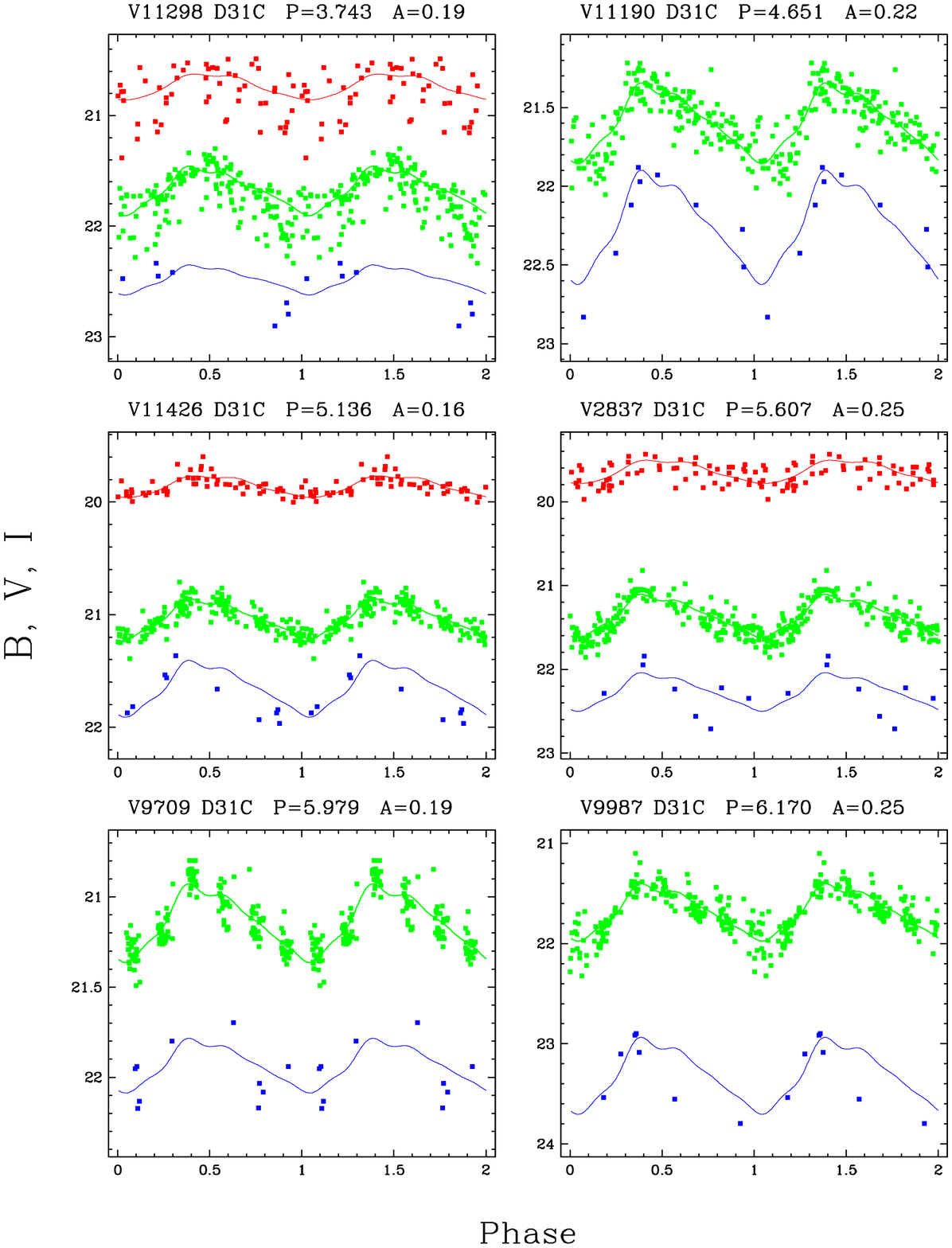}{19.5cm}{0}{83}{83}{-260}{-40}
\caption{$BVI$ lightcurves of Cepheid variables found in the field
M31C. The thin continuous line represents the best fit Cepheid
template for each star and photometric band. $B$ (if present) is
always the faintest and $I$ (if present) is always the brightest.}
\label{fig:ceph}
\end{figure}

\addtocounter{figure}{-1}
\begin{figure}[p]
\plotfiddle{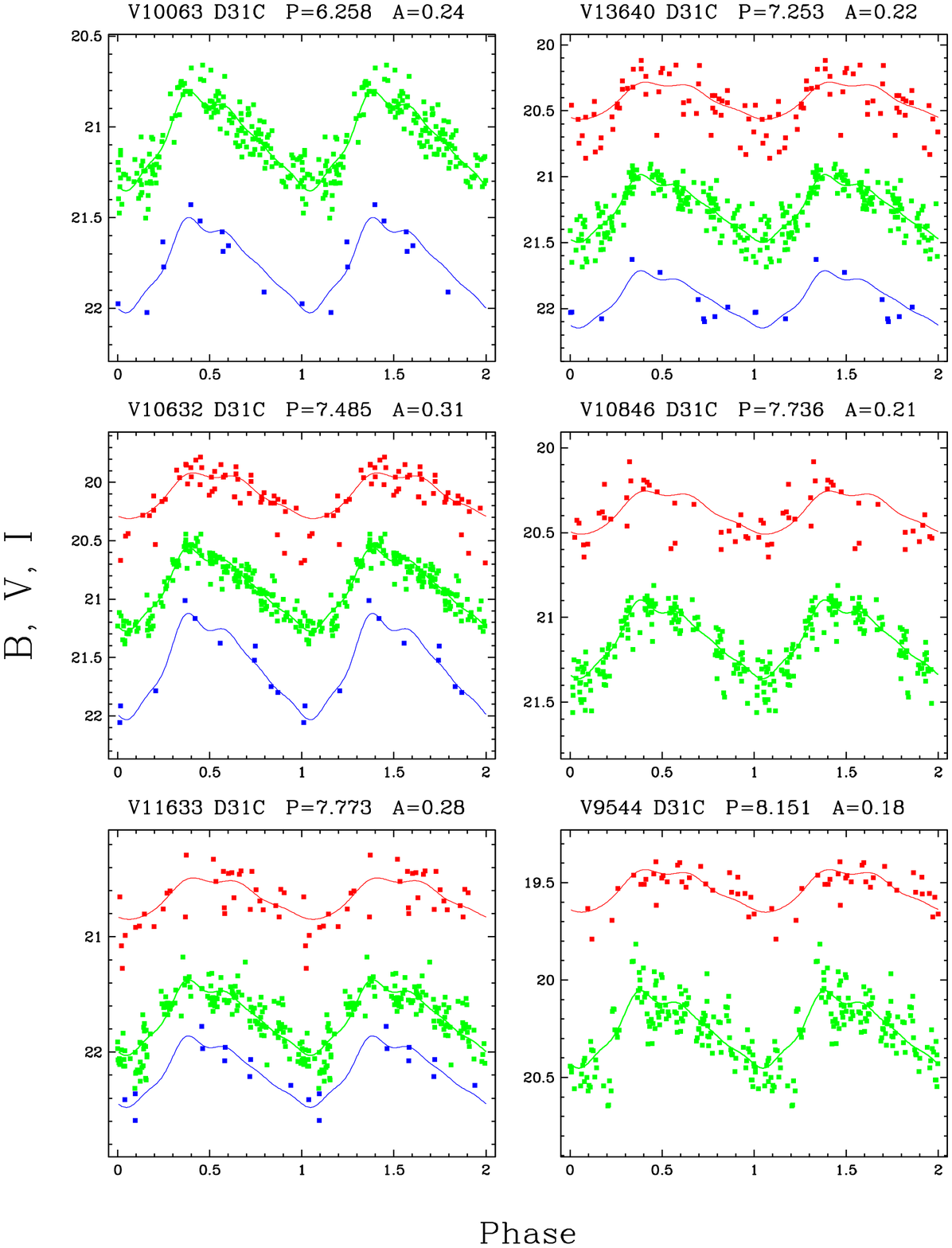}{19.5cm}{0}{83}{83}{-260}{-40}
\caption{Continued.}
\end{figure}

\addtocounter{figure}{-1}
\begin{figure}[p]
\plotfiddle{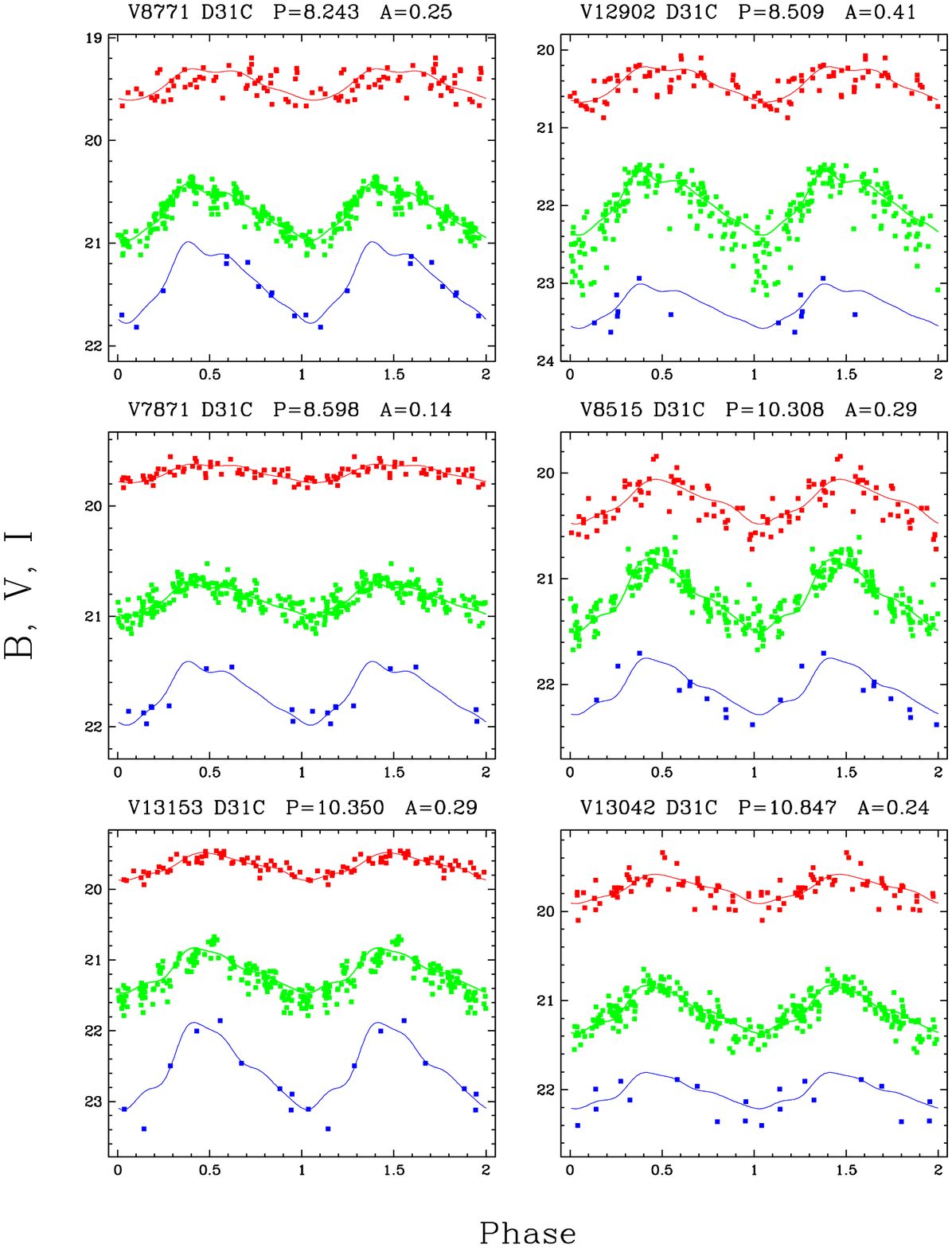}{19.5cm}{0}{83}{83}{-260}{-40}
\caption{Continued.}
\end{figure}

\addtocounter{figure}{-1}
\begin{figure}[p]
\plotfiddle{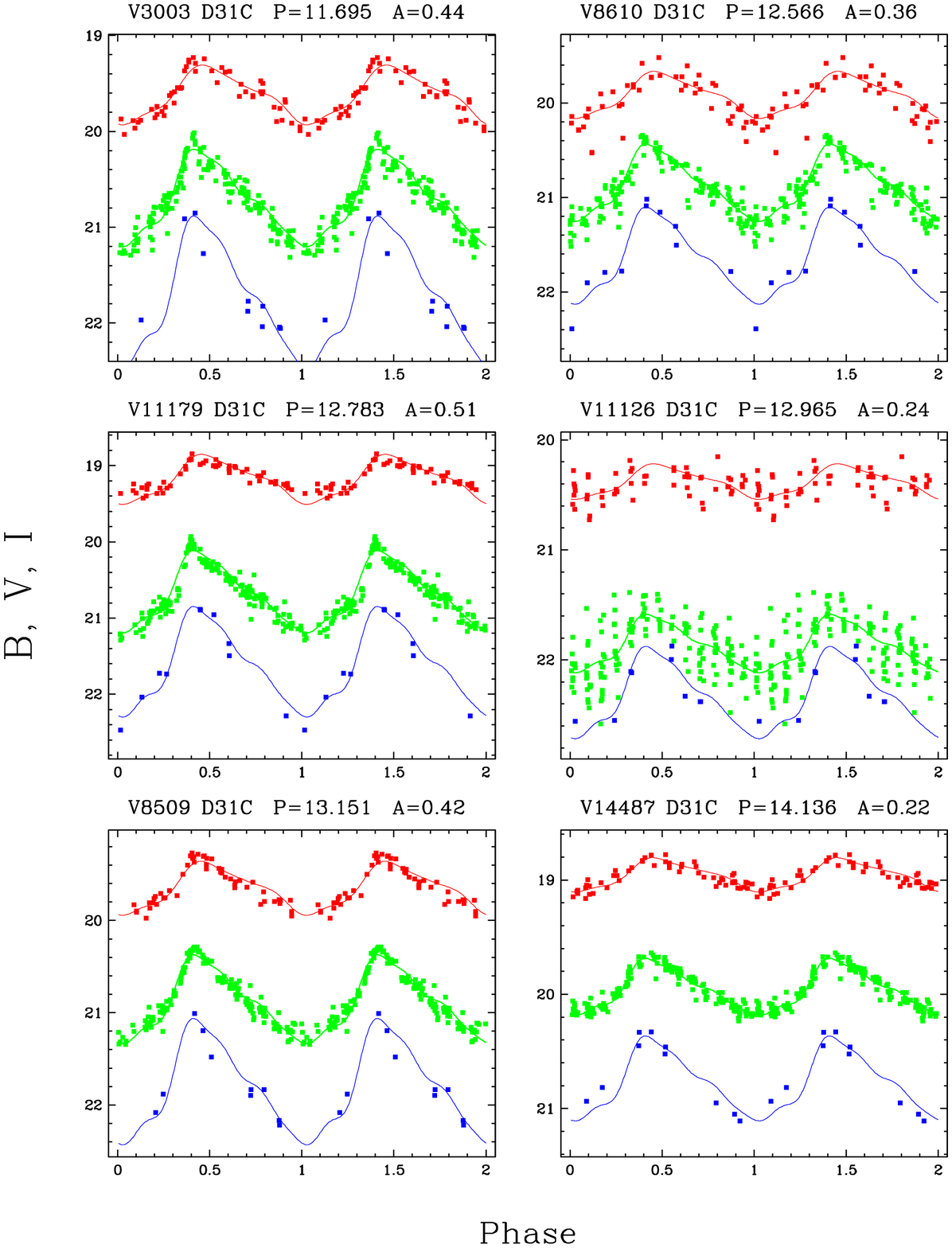}{19.5cm}{0}{83}{83}{-260}{-40}
\caption{Continued.}
\end{figure}

\addtocounter{figure}{-1}
\begin{figure}[p]
\plotfiddle{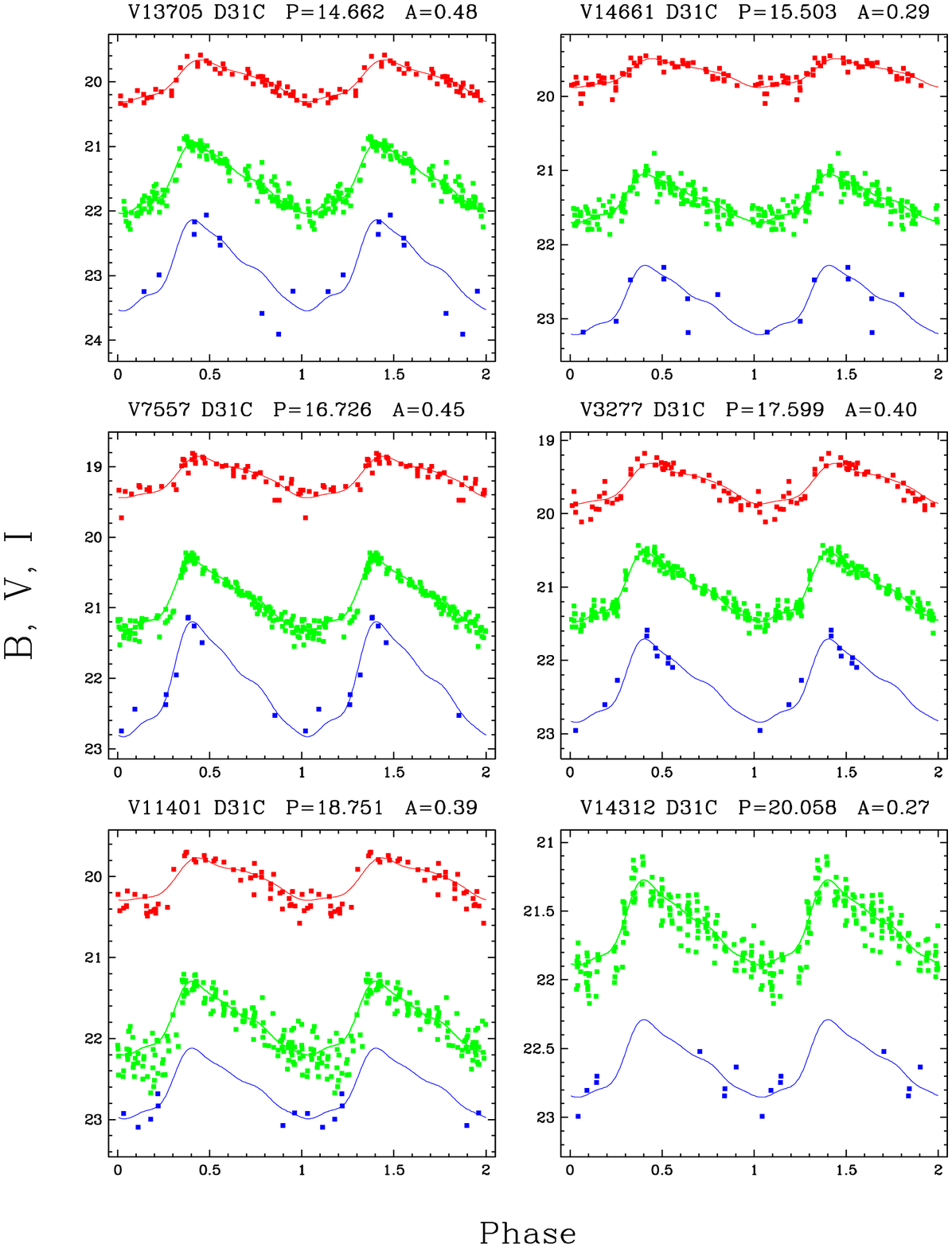}{19.5cm}{0}{83}{83}{-260}{-40}
\caption{Continued.}
\end{figure}

\addtocounter{figure}{-1}
\begin{figure}[p]
\plotfiddle{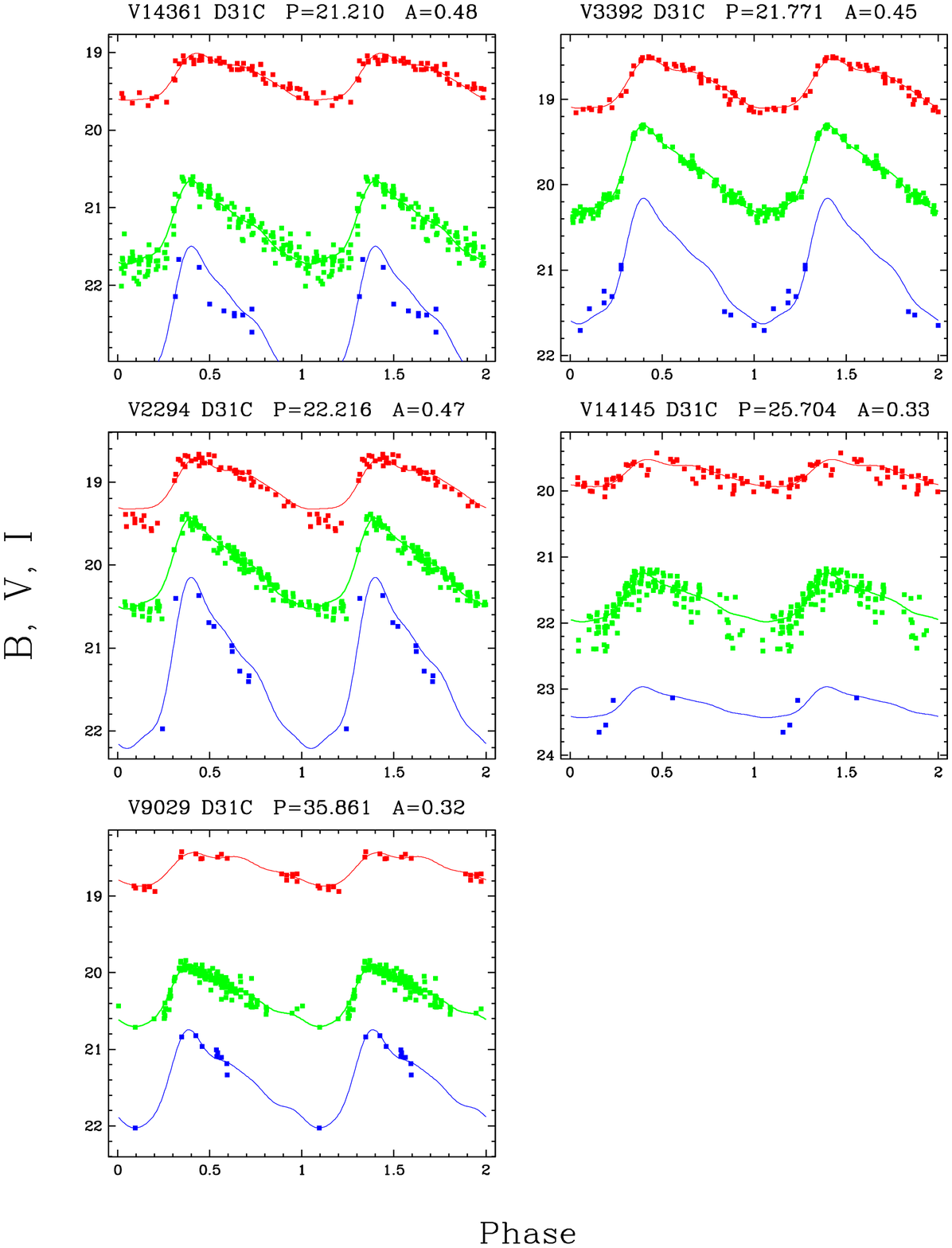}{19.5cm}{0}{83}{83}{-260}{-40}
\caption{Continued.}
\end{figure}

\subsection{Other periodic variables}

For many of the variables preliminary classified as Cepheids we
decided upon closer examination to classify them as ``other periodic
variables''.  In Table~\ref{table:per} we present the parameters of 37
possible periodic variables, other than Cepheids and eclipsing
binaries, in the M31C field, sorted by the increasing period $P$.  For
each variable we present its name, J2000.0 coordinates, period $P$,
error-weighted mean magnitudes $\bar{V}$ and (when available)
$\bar{I}, \bar{B}$. To quantify the amplitude of the variability, we
also give the standard deviations of the measurements in the $BVI$
bands, $\sigma_{V},\sigma_{I}$ and $\sigma_{B}$.

Note that in most cases the periods were derived by fitting the
template Cepheids lightcurves, so they should only be treated as the
first approximation of the true period. Many of these periodic
variables are Type II Cepheids (W Virginis and RV Tauri variables),
based on their light curves and their location on the P-L diagram
(Figure~\ref{fig:pl}).

\begin{figure}[p]
\plotfiddle{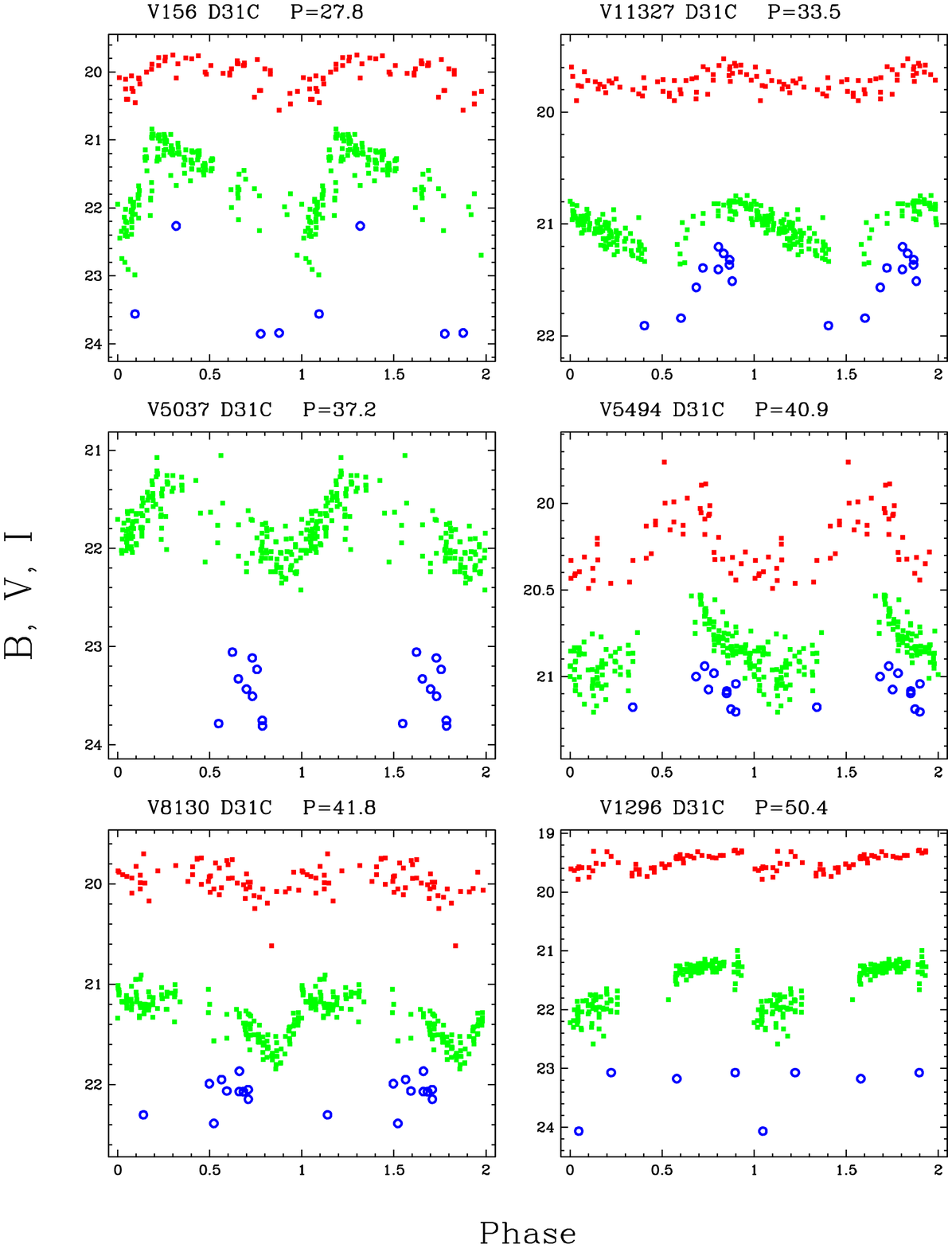}{19.5cm}{0}{83}{83}{-260}{-40}
\caption{$BVI$ lightcurves of other periodic variables found in the
field M31C.  $B$-band data (shown with the open circles, if present)
is usually the faintest and $I$ (if present) is usually the
brightest.}
\label{fig:per}
\end{figure}

\addtocounter{figure}{-1}
\begin{figure}[p]
\plotfiddle{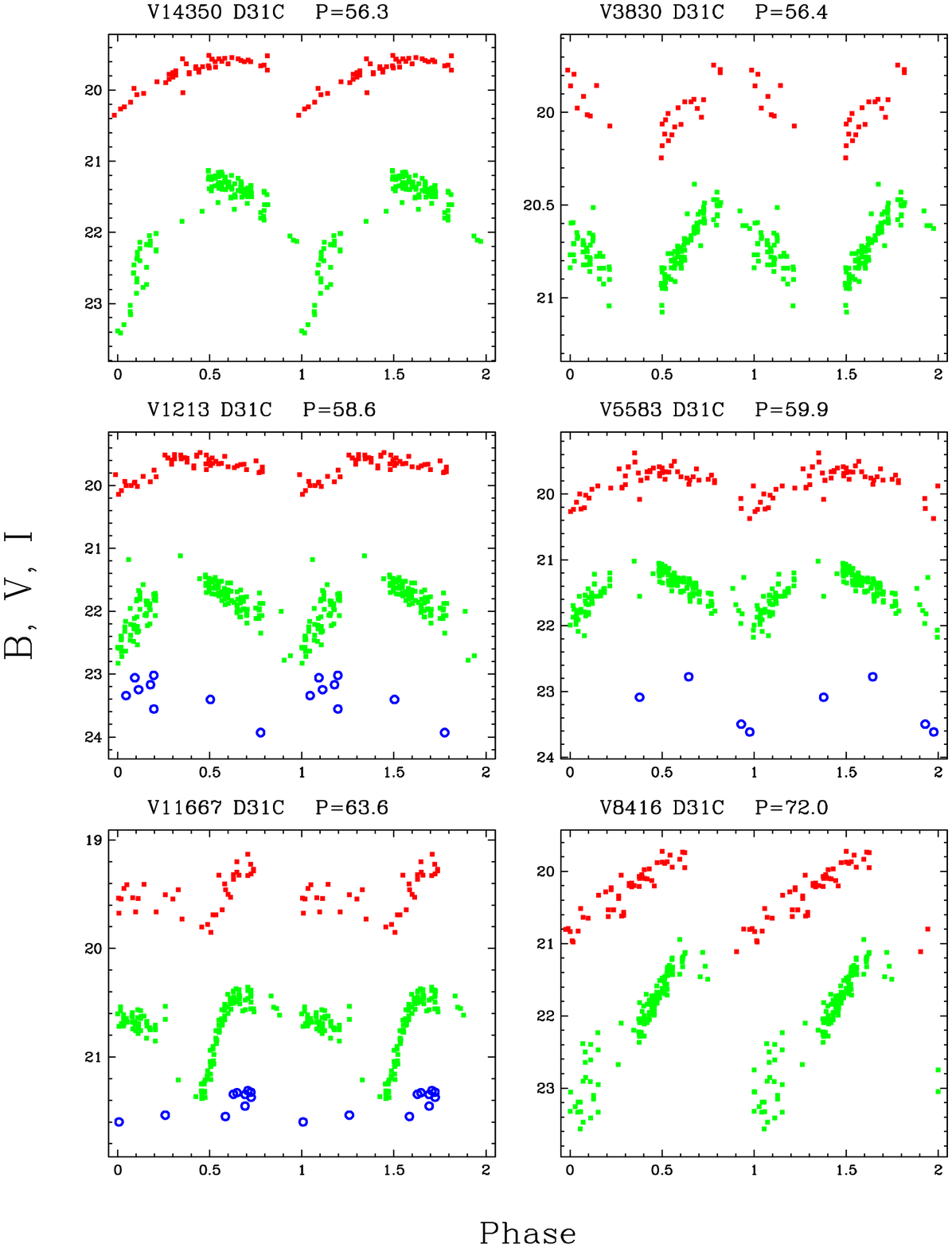}{19.5cm}{0}{83}{83}{-260}{-40}
\caption{Continued.}
\end{figure}

\addtocounter{figure}{-1}
\begin{figure}[p]
\plotfiddle{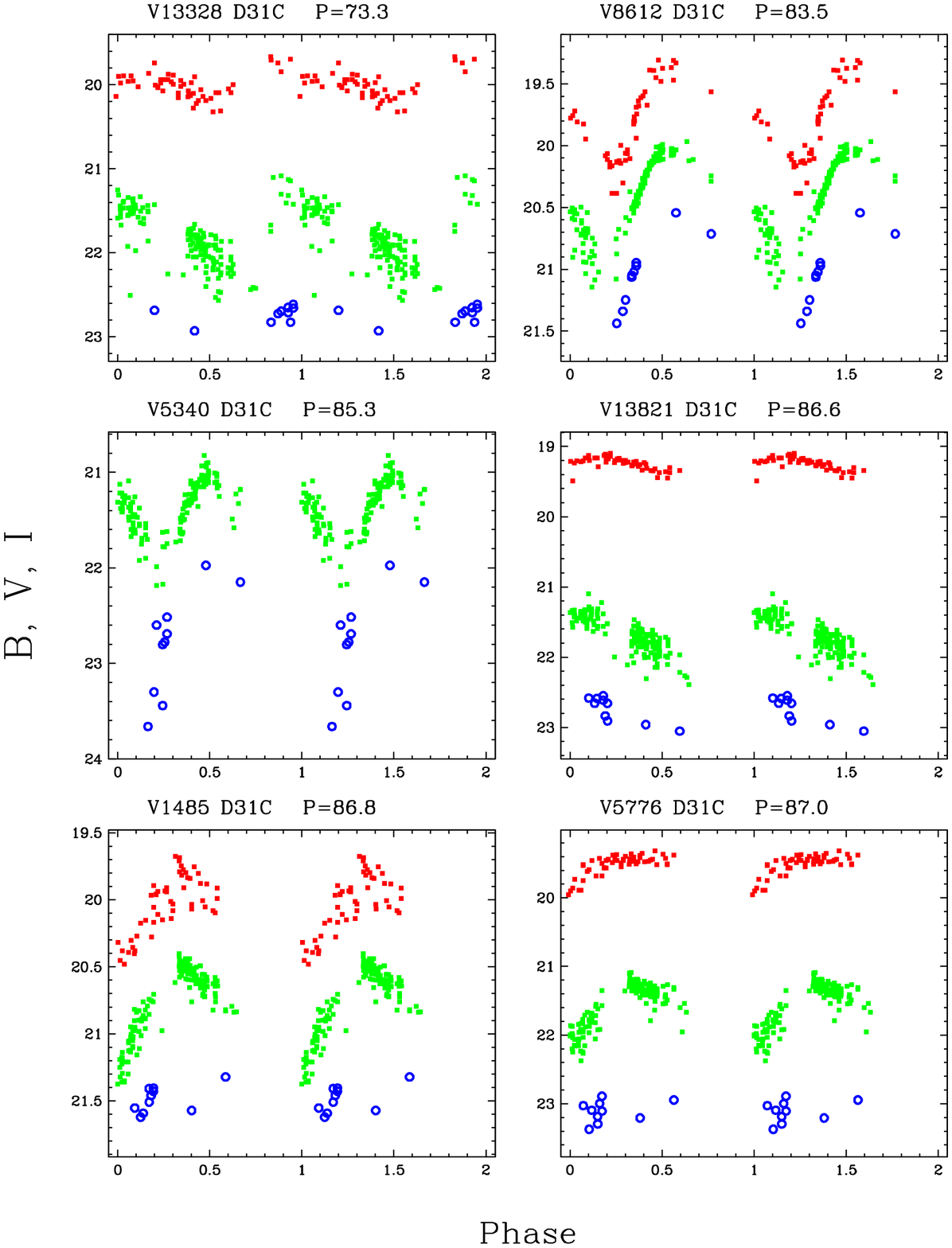}{19.5cm}{0}{83}{83}{-260}{-40}
\caption{Continued.}
\end{figure}

\addtocounter{figure}{-1}
\begin{figure}[p]
\plotfiddle{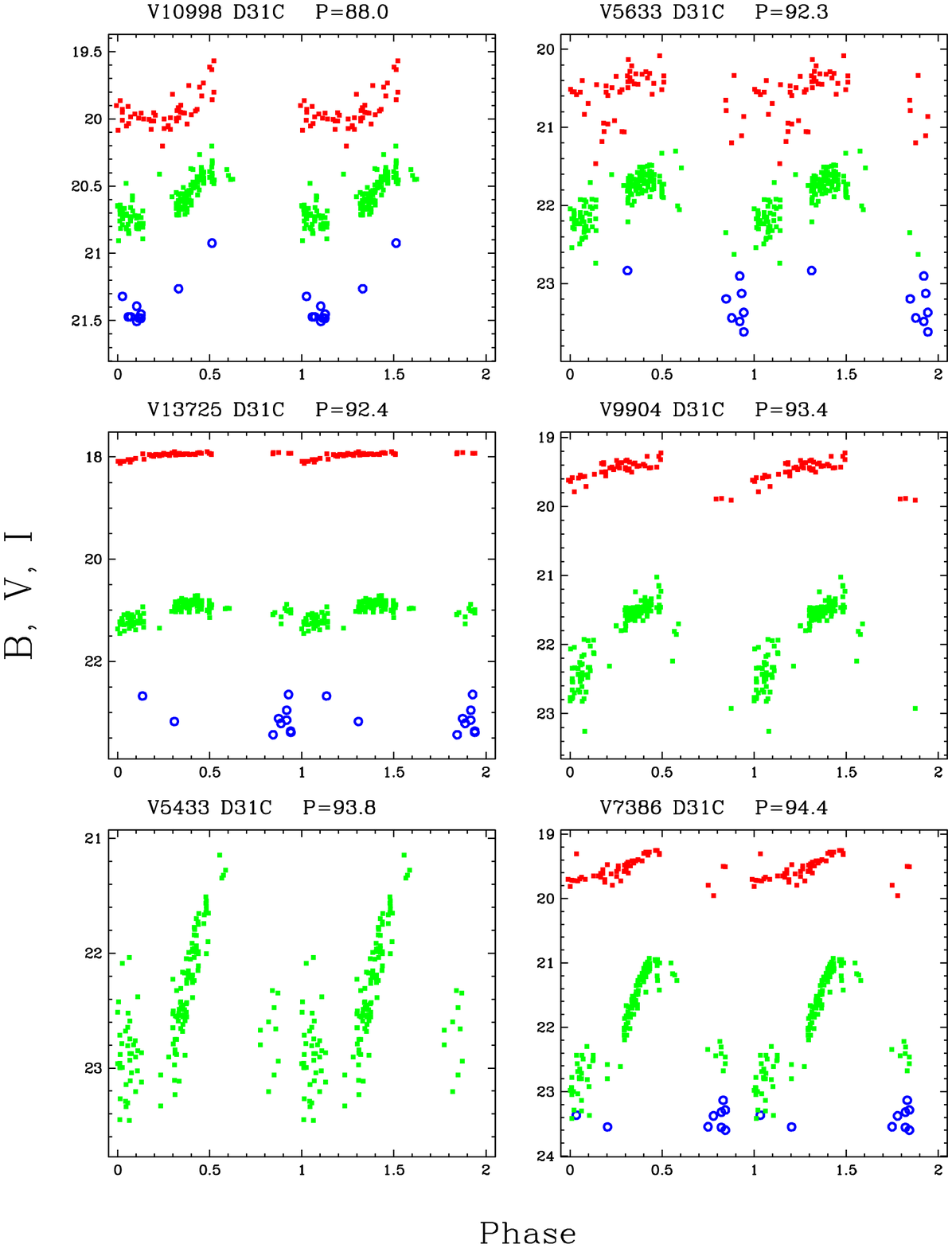}{19.5cm}{0}{83}{83}{-260}{-40}
\caption{Continued.}
\end{figure}

\addtocounter{figure}{-1}
\begin{figure}[p]
\plotfiddle{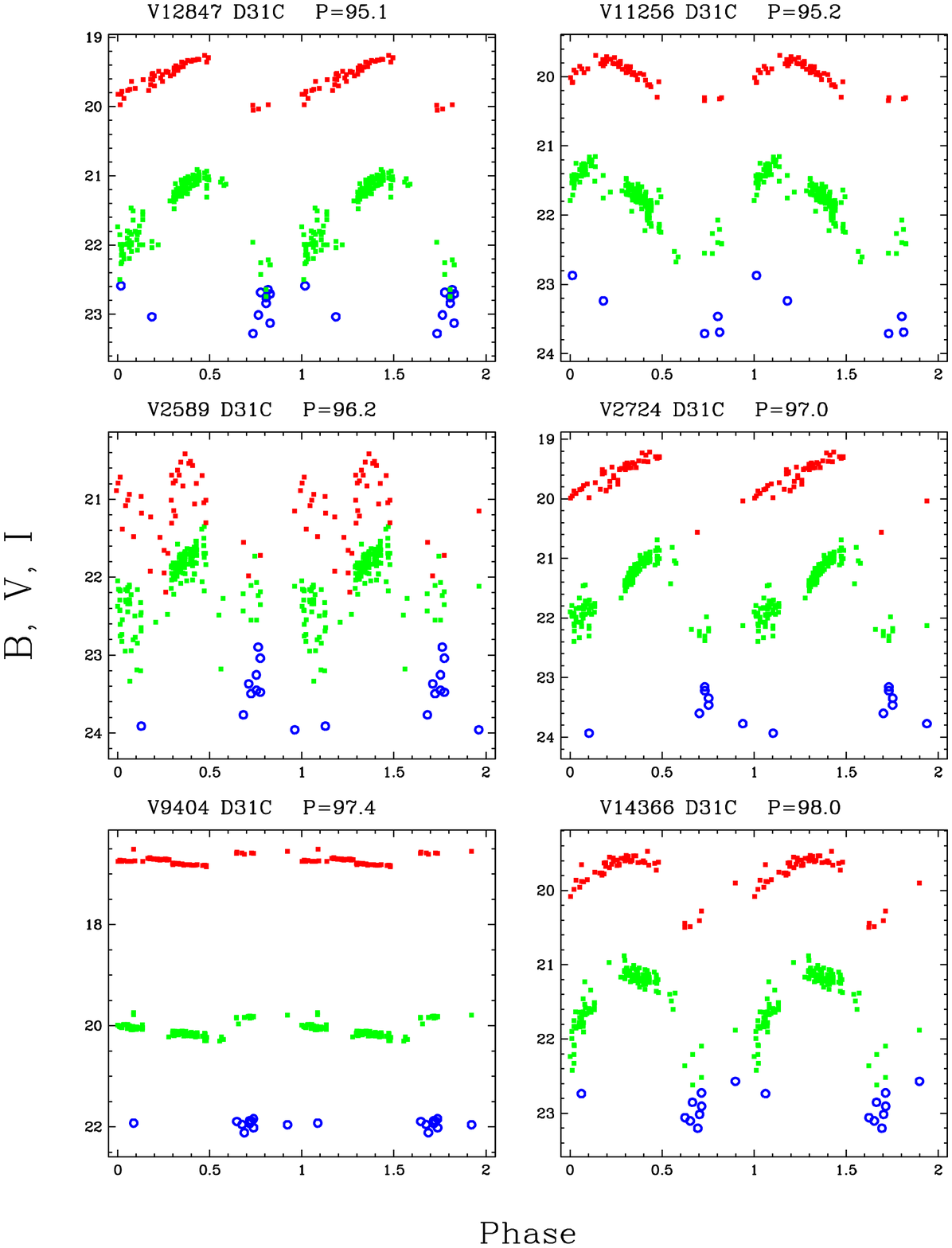}{19.5cm}{0}{83}{83}{-260}{-40}
\caption{Continued.}
\end{figure}

\addtocounter{figure}{-1}
\begin{figure}[p]
\plotfiddle{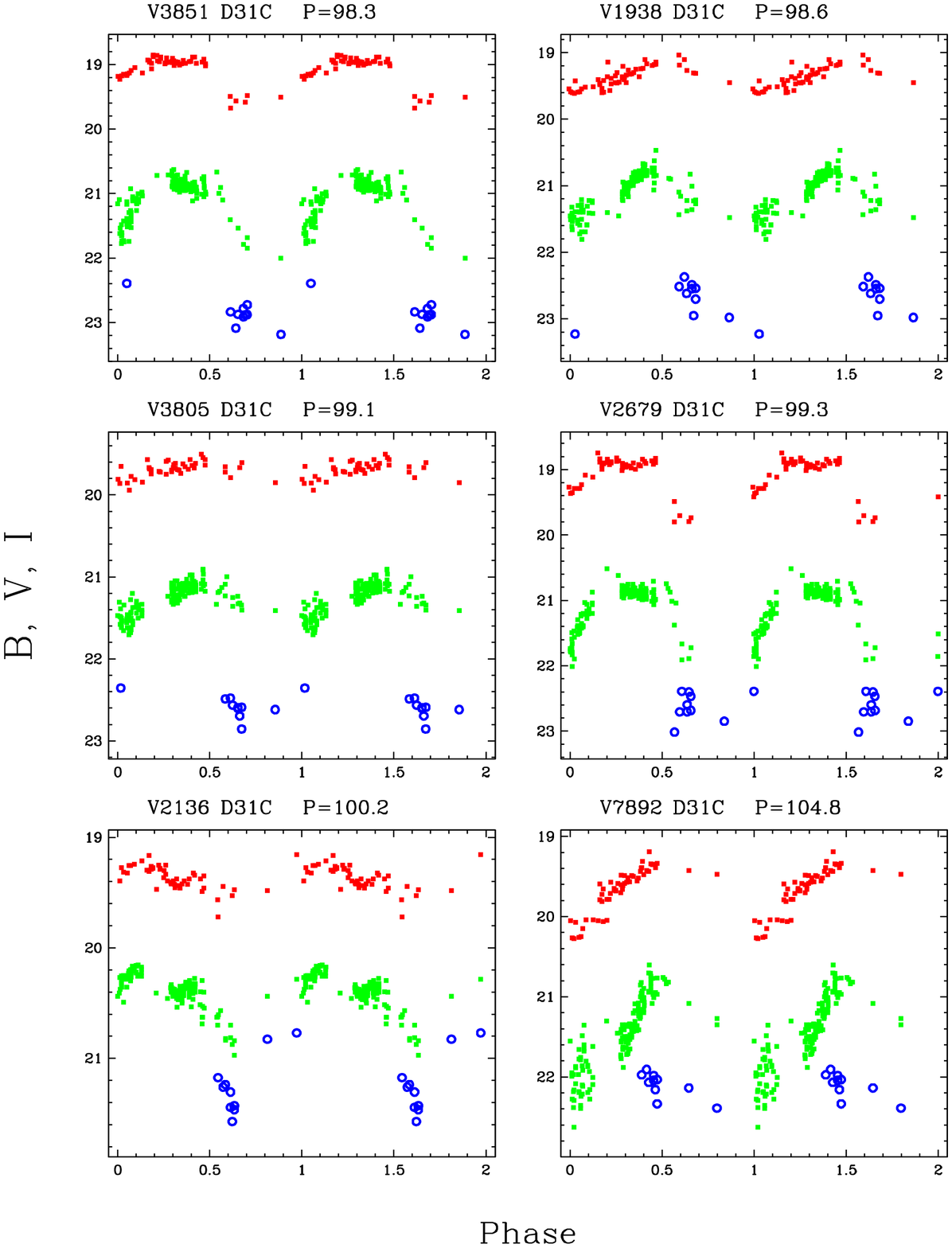}{19.5cm}{0}{83}{83}{-260}{-40}
\caption{Continued.}
\end{figure}

\addtocounter{figure}{-1}
\begin{figure}[p]
\plotfiddle{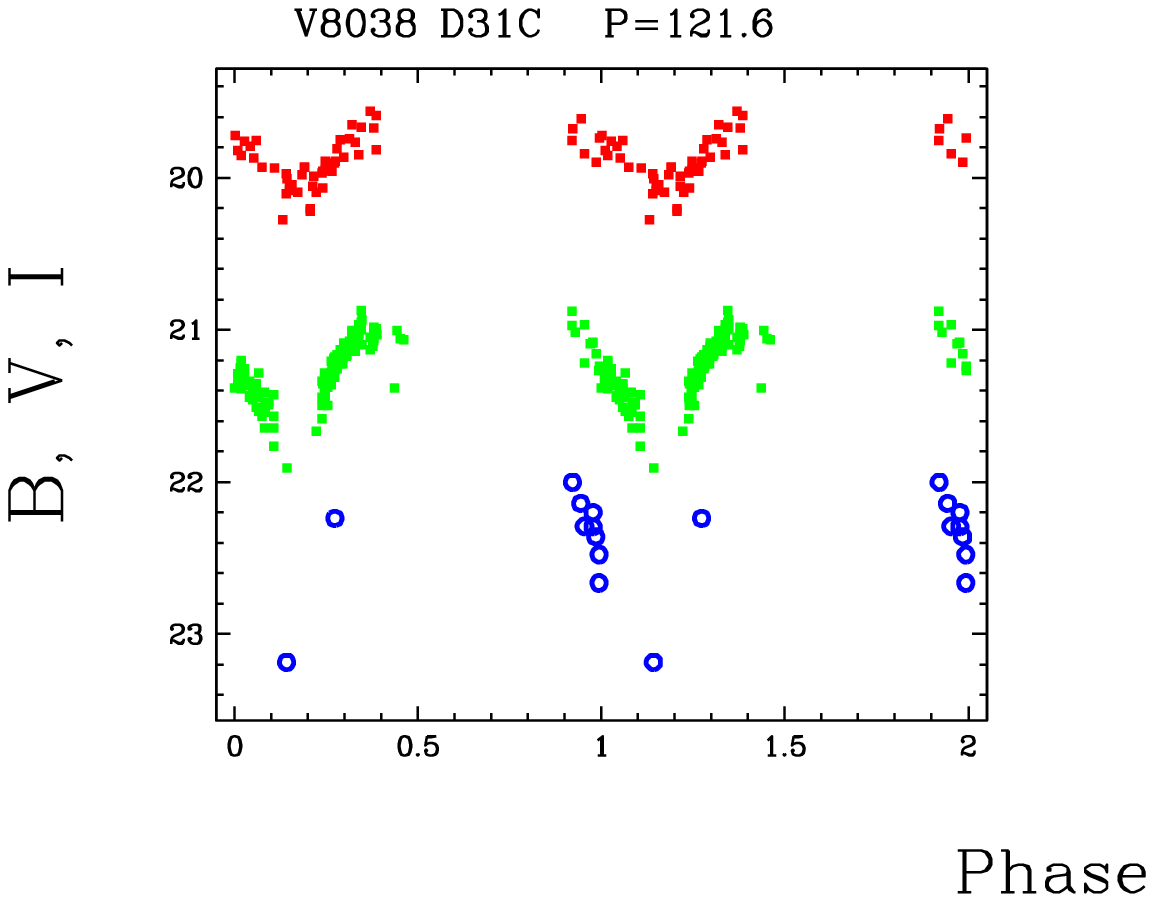}{4.75cm}{0}{83}{83}{-260}{-415}
\caption{Continued.}
\end{figure}

\begin{figure}[t]
\plotfiddle{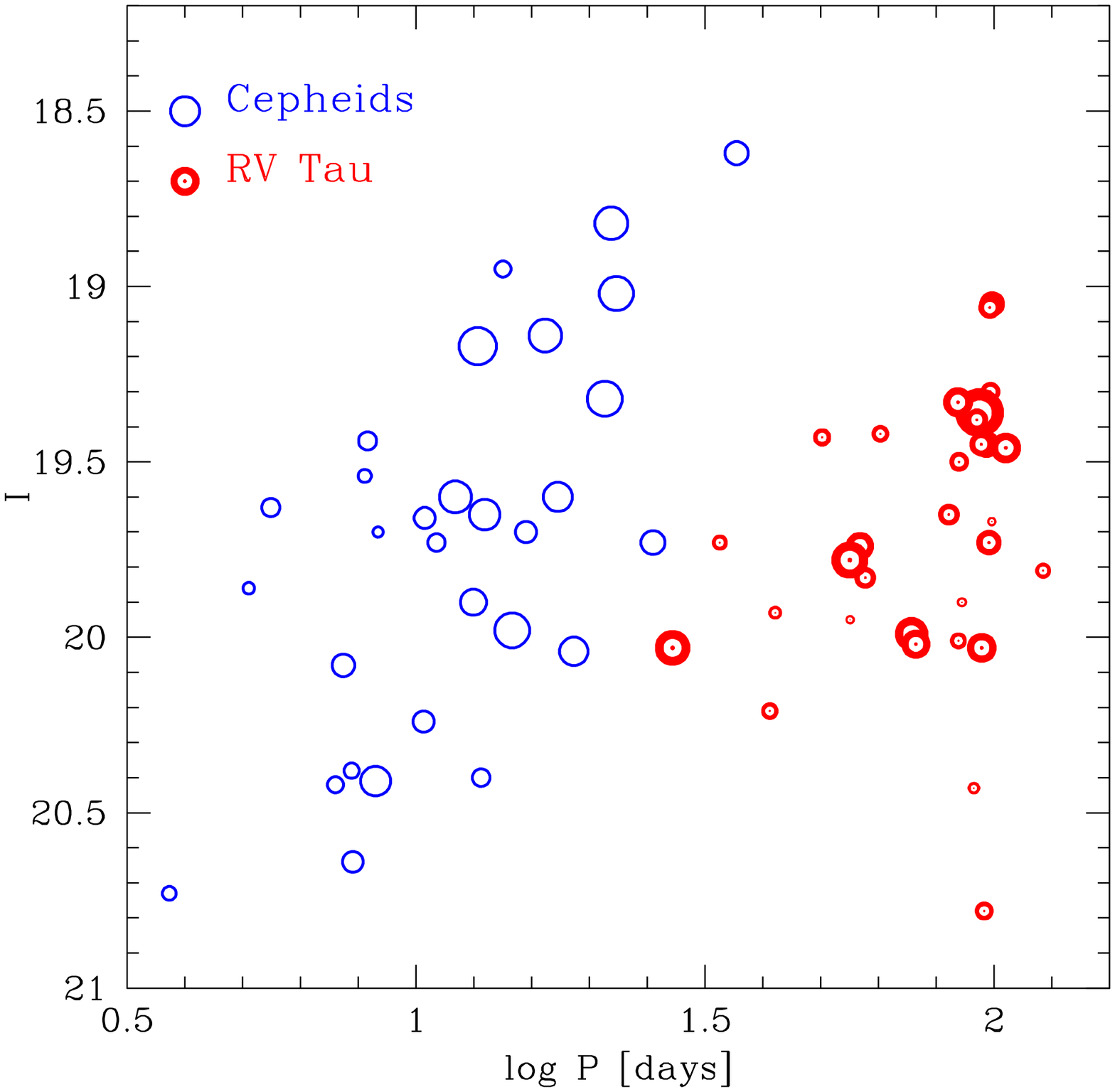}{8cm}{0}{50}{50}{-160}{-85}
\caption{Diagram of $\log{P}$ vs. $I$  for the
Cepheids (open circles) and RV Tau (dotted circles) variables. The
sizes of the circles are proportional to the $V$ amplitude of the
variability.}
\label{fig:pl}
\end{figure}

\begin{small}
\tablenum{2}
\begin{planotable}{lrrrrrrrl}
\tablewidth{35pc}
\tablecaption{DIRECT Cepheids in M31C}
\tablehead{ \colhead{Name} & \colhead{$\alpha_{J2000.0}$} &
\colhead{$\delta_{J2000.0}$} & \colhead{$P$}  &
\colhead{} & \colhead{} & \colhead{} & \colhead{} & \colhead{} \\ 
\colhead{(D31C)} & \colhead{(deg)} & \colhead{(deg)}
& \colhead{$(days)$} & \colhead{$\langle V\rangle$} & \colhead{$\langle I\rangle$} 
& \colhead{$\langle B\rangle$} & \colhead{$A$}  & \colhead{Comments} } 
\startdata 
V11298\ldots   & 11.1346 & 41.3806 &  3.743 & 21.67 & 20.73 & 22.48 & 0.19 & \nl
V11190\dotfill & 11.1339 & 41.3424 &  4.651 & 21.56 &\nodata& 22.22 & 0.22 & \nl
V11426\dotfill & 11.1363 & 41.4097 &  5.136 & 21.03 & 19.86 & 21.64 & 0.16 & Ma97 85\nl
V2837\dotfill  & 11.0159 & 41.4518 &  5.607 & 21.34 & 19.63 & 22.25 & 0.25 & \nl
V9709\dotfill  & 11.1081 & 41.4021 &  5.979 & 21.11 &\nodata& 21.93 & 0.19 & \nl
V9987\dotfill  & 11.1141 & 41.3523 &  6.170 & 21.62 &\nodata& 23.28 & 0.25 & \nl
V10063\dotfill & 11.1129 & 41.4347 &  6.258 & 21.04 &\nodata& 21.74 & 0.24 & \nl
V13640\dotfill & 11.1831 & 41.4072 &  7.253 & 21.23 & 20.42 & 21.92 & 0.22 & \nl
V10632\dotfill & 11.1225 & 41.4122 &  7.485 & 20.88 & 20.08 & 21.54 & 0.31 & Ma97 83\nl
V10846\dotfill & 11.1229 & 41.5087 &  7.736 & 21.12 & 20.38 &\nodata& 0.21 & V1562 D31B\nl
V11633\dotfill & 11.1414 & 41.3671 &  7.773 & 21.67 & 20.64 & 22.16 & 0.28 &  \nl
V9544\dotfill  & 11.1021 & 41.5129 &  8.151 & 20.25 & 19.54 &\nodata& 0.18 & V643 D31B \nl
V8771\dotfill  & 11.0909 & 41.4971 &  8.243 & 20.68 & 19.44 & 21.35 & 0.25 & V129 D31B \nl
V12902\dotfill & 11.1635 & 41.5022 &  8.509 & 21.93 & 20.41 & 23.28 & 0.41 & V2977 D31B\nl
V7871\dotfill  & 11.0829 & 41.3443 &  8.598 & 20.84 & 19.70 & 21.67 & 0.14 & \nl
V8515\dotfill  & 11.0915 & 41.3549 & 10.308 & 21.16 & 20.24 & 22.02 & 0.29 & \nl
V13153\dotfill & 11.1715 & 41.4072 & 10.350 & 21.14 & 19.66 & 22.46 & 0.29 & Ma97 90\nl
V13042\dotfill & 11.1698 & 41.3994 & 10.847 & 21.10 & 19.73 & 22.01 & 0.24 & \nl
V3003\dotfill  & 11.0171 & 41.4620 & 11.695 & 20.68 & 19.60 & 21.61 & 0.44 & \nl
V8610\dotfill  & 11.0916 & 41.3966 & 12.566 & 20.85 & 19.90 & 21.61 & 0.36 & Ma97 77\nl
V11179\dotfill & 11.1317 & 41.4136 & 12.783 & 20.64 & 19.17 & 21.53 & 0.51 & \nl
V11126\dotfill & 11.1299 & 41.4356 & 12.965 & 21.85 & 20.40 & 22.30 & 0.24 & \nl
V8509\dotfill  & 11.0909 & 41.3723 & 13.151 & 20.85 & 19.65 & 21.72 & 0.42 & Ma97 76\nl
V14487\dotfill & 11.2027 & 41.4876 & 14.136 & 19.95 & 18.95 & 20.75 & 0.22 & Ma97 95\nl
V13705\dotfill & 11.1822 & 41.4766 & 14.662 & 21.50 & 19.98 & 22.83 & 0.48 & \nl
V14661\dotfill & 11.2092 & 41.4527 & 15.503 & 21.39 & 19.70 & 22.76 & 0.29 & \nl
V7557\dotfill  & 11.0783 & 41.3467 & 16.726 & 20.85 & 19.14 & 21.98 & 0.45 & Ma97 75\nl
V3277\dotfill  & 11.0238 & 41.3485 & 17.599 & 21.02 & 19.60 & 22.29 & 0.40 & \nl
V11401\dotfill & 11.1358 & 41.4061 & 18.751 & 21.77 & 20.04 & 22.59 & 0.39 & \nl
V14312\dotfill & 11.2015 & 41.3861 & 20.058 & 21.57 &\nodata& 22.57 & 0.27 & \nl
V14361\dotfill & 11.2006 & 41.4493 & 21.210 & 21.21 & 19.32 & 22.36 & 0.48 & Ma97 94\nl
V3392\dotfill  & 11.0229 & 41.4243 & 21.771 & 19.85 & 18.82 & 20.91 & 0.45 & \nl
V2294\dotfill  & 11.0078 & 41.5020 & 22.216 & 20.01 & 19.02 & 21.15 & 0.47 & \nl
V14145\dotfill & 11.1950 & 41.4465 & 25.704 & 21.64 & 19.73 & 23.22 & 0.33 & \nl
V9029\dotfill  & 11.0996 & 41.3533 & 35.861 & 20.31 & 18.62 & 21.39 & 0.32 & 
\enddata
\label{table:ceph}
\tablecomments{V10846 D31C was found in Paper I as V1562 D31B, with
$P=7.784\;days$, $\langle V\rangle = 21.20$ and $\langle I\rangle =
20.43$; V9544 D31C was found as V643 D31B, with $P=7.889\;days$,
$\langle V\rangle = 20.39$ and $\langle I\rangle = 19.52$; V8771 D31C
was found as V129 D31B, with $P=8.242\;days$, $\langle V\rangle =
20.74$ and $\langle I\rangle = 19.58$; V12902 D31C was found as V2977
D31B, with $P=8.518\;days$, $\langle V\rangle = 21.80$ and $\langle
I\rangle = 20.40$.}
\end{planotable}
\end{small}

\subsection{Miscellaneous  variables}	
	
In Table~\ref{table:misc} we present the parameters of 31
miscellaneous variables in the M31C field, sorted by increasing value
of the mean magnitude $\bar{V}$. For each variable we present its
name, J2000.0 coordinates and mean magnitudes $\bar{V}, \bar{I}$ and
$\bar{B}$.  To quantify the amplitude of the variability, we also give
the standard deviations of the measurements in $VIB$ bands,
$\sigma_{V}, \sigma_{I}$ and $\sigma_{B}$.  In the ``Comments'' column
we give a rather broad sub-classification of the variability: LP --
possible long-period variable; Irr -- irregular variable.  In
Figure~\ref{fig:misc} we show the unphased $VI$ lightcurves of the
miscellaneous variables.

Most of the miscellaneous variables seem to represent the LP type of
variability, with few variables showing irregular variations. However,
inspection of the color-magnitude diagram (Figure~\ref{fig:cmd})
reveals that many of the miscellaneous variables land in the CMD in
the same area as the RV Tau variables, which suggests they are Type II
Cepheids.

\begin{small}
\tablenum{3} 
\begin{planotable}{cccrccccccl}
\tablewidth{40pc}
\tablecaption{DIRECT Other Periodic Variables in M31C}
\tablehead{ \colhead{Name} & \colhead{$\alpha_{J2000.0}$} &
\colhead{$\delta_{J2000.0}$} & \colhead{$P$} &
\colhead{} & \colhead{} & \colhead{} & \colhead{} & \colhead{} & \colhead{} & \colhead{} \\
\colhead{(D31C)} &  \colhead{(deg)} &  \colhead{(deg)} & 
\colhead{$(days)$} & \colhead{$\bar{V}$} &
\colhead{$\bar{I}$} & \colhead{$\bar{B}$} & \colhead{$\sigma_V$} & 
\colhead{$\sigma_I$} & \colhead{$\sigma_B$} & \colhead{Comments} }
\startdata     
V156\dotfill   & 10.9877 & 41.3845 &  27.8 & 21.18 & 20.00 & 22.45 & 0.49 & 0.22 & 0.76 & RV Tau \nl
V11327\dotfill & 11.1349 & 41.3930 &  33.5 & 20.97 & 19.71 & 21.43 & 0.15 & 0.09 & 0.24 & \nl
V5037\dotfill  & 11.0424 & 41.4547 &  37.2 & 21.72 &\nodata& 23.36 & 0.30 &\nodata& 0.29 & \nl
V5494\dotfill  & 11.0515 & 41.3655 &  40.9 & 20.81 & 20.17 & 21.06 & 0.15 & 0.18 & 0.09 & \nl
V8130\dotfill  & 11.0829 & 41.4698 &  41.8 & 21.31 & 19.94 & 22.06 & 0.22 & 0.15 & 0.16 & \nl
V1296\dotfill  & 11.0011 & 41.3400 &  50.4 & 21.36 & 19.46 & 23.15 & 0.37 & 0.13 & 0.48 & \nl  
V14350\ldots   & 11.2033 & 41.3398 &  56.3 & 21.36 & 19.71 &\nodata& 0.55 & 0.21 &\nodata& \nl
V3830\dotfill  & 11.0256 & 41.5107 &  56.4 & 20.72 & 19.96 &\nodata& 0.14 & 0.13 &\nodata& \nl
V1213\dotfill  & 10.9989 & 41.3831 &  58.6 & 21.73 & 19.66 & 23.24 & 0.33 & 0.17 & 0.30 & \nl
V5583\dotfill  & 11.0529 & 41.3526 &  59.9 & 21.37 & 19.75 & 23.04 & 0.24 & 0.22 & 0.39 & \nl
V11667\dotfill & 11.1391 & 41.4679 &  63.6 & 20.63 & 19.42 & 21.41 & 0.27 & 0.18 & 0.11 & RV Tau\nl
V8416\dotfill  & 11.0869 & 41.4684 &  72.0 & 21.62 & 20.09 &\nodata& 0.62 & 0.37 &\nodata& \nl  
V13328\dotfill & 11.1725 & 41.4973 &  73.3 & 21.67 & 19.97 & 22.73 & 0.33 & 0.15 & 0.10 & \nl
V8612\dotfill  & 11.0922 & 41.3798 &  83.5 & 20.24 & 19.70 & 20.89 & 0.29 & 0.31 & 0.27 & RV Tau\nl
V5340\dotfill  & 11.0475 & 41.4341 &  85.3 & 21.21 &\nodata& 22.33 & 0.25 &\nodata& 0.54 & \nl
V13821\dotfill & 11.1879 & 41.3777 &  86.6 & 21.62 & 19.23 & 22.71 & 0.25 & 0.09 & 0.18 & \nl
V1485\dotfill  & 11.0030 & 41.3578 &  86.8 & 20.61 & 19.96 & 21.46 & 0.25 & 0.22 & 0.10 & \nl
V5776\dotfill  & 11.0518 & 41.4741 &  87.0 & 21.38 & 19.48 & 23.07 & 0.32 & 0.15 & 0.15 & \nl  
V10998\dotfill & 11.1257 & 41.4913 &  88.0 & 20.56 & 19.93 & 21.33 & 0.15 & 0.12 & 0.18 & \nl
V5633\dotfill  & 11.0524 & 41.3904 &  92.3 & 21.75 & 20.45 & 23.10 & 0.28 & 0.30 & 0.28 & \nl
V13725\dotfill & 11.1842 & 41.4296 &  92.4 & 20.93 & 17.97 & 22.98 & 0.17 & 0.05 & 0.28 & LP \nl
V9904\dotfill  & 11.1114 & 41.3963 &  93.4 & 21.55 & 19.45 &\nodata& 0.48 & 0.15 &\nodata& \nl
V5433\dotfill  & 11.0498 & 41.3979 &  93.8 & 22.10 &\nodata&\nodata& 0.51 &\nodata&\nodata& \nl
V7386\dotfill  & 11.0716 & 41.4897 &  94.4 & 21.33 & 19.49 & 23.38 & 0.69 & 0.17 & 0.16 & \nl  
V12847\dotfill & 11.1635 & 41.4683 &  95.1 & 21.18 & 19.55 & 22.80 & 0.45 & 0.21 & 0.23 & \nl
V11256\dotfill & 11.1332 & 41.4051 &  95.2 & 21.57 & 19.91 & 23.20 & 0.32 & 0.17 & 0.35 & \nl
V2589\dotfill  & 11.0152 & 41.3759 &  96.2 & 21.84 & 20.81 & 23.27 & 0.39 & 0.45 & 0.35 & \nl
V2724\dotfill  & 11.0175 & 41.3498 &  97.0 & 21.19 & 19.55 & 23.44 & 0.44 & 0.26 & 0.29 & \nl
V9404\dotfill  & 11.1038 & 41.3895 &  97.4 & 20.08 & 16.75 & 21.93 & 0.11 & 0.08 & 0.08 & LP \nl
V14366\dotfill & 11.2033 & 41.3553 &  98.0 & 21.21 & 19.69 & 22.82 & 0.36 & 0.26 & 0.21 & \nl  
V3851\dotfill  & 11.0285 & 41.4277 &  98.3 & 20.91 & 18.99 & 22.73 & 0.30 & 0.20 & 0.21 & \nl
V1938\dotfill  & 11.0045 & 41.4926 &  98.6 & 20.95 & 19.32 & 22.61 & 0.28 & 0.15 & 0.27 & \nl
V3805\dotfill  & 11.0294 & 41.3836 &  99.1 & 21.20 & 19.68 & 22.53 & 0.18 & 0.10 & 0.13 & \nl
V2679\dotfill  & 11.0174 & 41.3357 &  99.3 & 20.92 & 18.97 & 22.54 & 0.32 & 0.27 & 0.21 & \nl
V2136\dotfill  & 11.0081 & 41.4458 & 100.2 & 20.35 & 19.34 & 21.09 & 0.15 & 0.11 & 0.27 & \nl
V7892\dotfill  & 11.0818 & 41.3915 & 104.8 & 21.22 & 19.56 & 22.09 & 0.44 & 0.28 & 0.16 & \nl  
V8038\dotfill  & 11.0851 & 41.3495 & 121.6 & 21.18 & 19.85 & 22.25 & 0.19 & 0.16 & 0.33 &
\enddata
\label{table:per}
\end{planotable}
\end{small}

\begin{small}
\tablenum{4} 
\begin{planotable}{llllllllll}
\tablewidth{35pc}
\tablecaption{DIRECT Miscellaneous Variables in M31C}
\tablehead{ \colhead{Name} & \colhead{$\alpha_{J2000.0}$} &
\colhead{$\delta_{J2000.0}$} & \colhead{} & \colhead{} &
\colhead{} & \colhead{} & \colhead{} & \colhead{} & \colhead{} \\
\colhead{(D31C)} & \colhead{(deg)} & \colhead{(deg)} &
\colhead{$\bar{V}$} & \colhead{$\bar{I}$} & \colhead{$\bar{B}$} &
\colhead{$\sigma_V$} & \colhead{$\sigma_I$} & \colhead{$\sigma_B$} & \colhead{Comments} }
\startdata
V5497\dotfill  &  11.0489 & 41.4535 & 16.25 & 15.33 & 17.62 &  0.07 &  0.05 &  0.02 &   \nl
V9306\dotfill  &  11.1039 & 41.3451 & 16.60 & 15.98 & 16.97 &  0.04 &  0.03 &  0.02 &   \nl
V11839\ldots   &  11.1443 & 41.4177 & 16.74 & 16.61 & 16.65 &  0.04 &  0.03 &  0.01 &   \nl
V9102\dotfill  &  11.1007 & 41.3543 & 16.74 & 15.64 & 17.49 &  0.04 &  0.03 &  0.04 &   \nl 
V12381\dotfill &  11.1542 & 41.4424 & 17.12 & 16.52 & 17.54 &  0.04 &  0.02 &  0.03 &   \nl
V13814\dotfill &  11.1854 & 41.4551 & 18.07 & 17.65 & 18.34 &  0.05 &  0.03 &  0.03 &   \nl
V13102\dotfill &  11.1691 & 41.4512 & 18.16 & 16.77 & 19.52 &  0.07 &  0.03 &  0.04 &   \nl
V13833\dotfill &  11.1854 & 41.4678 & 18.25 & 16.67 & 19.09 &  0.05 &  0.04 &  0.03 &   \nl
V9360\dotfill  &  11.1033 & 41.3911 & 18.84 & 16.72 & 20.50 &  0.04 &  0.04 &  0.05 &   \nl
V10916\dotfill &  11.1269 & 41.4099 & 19.13 & 16.78 & 22.44 &  0.12 &  0.19 &  0.15 &   \nl
V11413\dotfill &  11.1367 & 41.3870 & 19.39 & 18.54 & 19.98 &  0.07 &  0.07 &  0.08 &   \nl
V8969\dotfill  &  11.0983 & 41.3713 & 19.63 & 17.13 & 22.20 &  0.07 &  0.02 &  0.10 &   \nl
V14370\dotfill &  11.1986 & 41.5138 & 19.72 & 16.69 &\nodata&  0.12 &  0.06 &\nodata& V4062 D31B  \nl
V13441\dotfill &  11.1767 & 41.4471 & 19.94 & 16.53 & 21.00 &  0.23 &  0.20 &  0.06 &   \nl
V5768\dotfill  &  11.0547 & 41.3761 & 20.03 & 19.40 & 20.39 &  0.09 &  0.08 &  0.04 &   \nl
V6175\dotfill  &  11.0605 & 41.3568 & 20.14 & 17.32 & 22.67 &  0.12 &  0.04 &  0.14 &   \nl
V9115\dotfill  &  11.1004 & 41.3696 & 20.23 & 17.77 & 22.05 &  0.10 &  0.03 &  0.11 &   \nl
V5588\dotfill  &  11.0488 & 41.4869 & 20.95 & 19.66 & 21.64 &  0.19 &  0.11 &  0.32 &   \nl
V5283\dotfill  &  11.0471 & 41.4166 & 21.02 & 19.01 & 22.30 &  0.41 &  0.11 &  0.17 &   \nl
V433\dotfill   &  10.9907 & 41.3868 & 21.07 & 19.06 & 22.56 &  0.24 &  0.26 &  0.13 &   \nl
V9205\dotfill  &  11.0987 & 41.4678 & 21.15 & 19.30 & 23.42 &  0.74 &  0.48 &  0.36 &   \nl
V499\dotfill   &  10.9906 & 41.4050 & 21.16 & 19.53 & 22.33 &  0.24 &  0.17 &  0.11 &   \nl
V14148\dotfill &  11.1968 & 41.3900 & 21.16 & 17.93 &\nodata&  0.37 &  0.40 &\nodata&   \nl
V7581\dotfill  &  11.0783 & 41.3585 & 21.26 &\nodata& 22.24 &  0.45 &\nodata&  0.41 &   \nl
V3225\dotfill  &  11.0230 & 41.3558 & 21.30 & 19.27 & 23.49 &  0.32 &  0.17 &  0.17 &   \nl
V4241\dotfill  &  11.0321 & 41.4616 & 21.66 & 18.41 &\nodata&  0.34 &  0.09 &\nodata&   \nl
V8170\dotfill  &  11.0830 & 41.4826 & 21.71 &\nodata& 23.18 &  0.34 &\nodata&  0.39 &   \nl
V11642\dotfill &  11.1412 & 41.3844 & 21.74 & 19.89 & 23.04 &  0.31 &  0.12 &  0.35 &   \nl 
V4674\dotfill  &  11.0385 & 41.4262 & 21.83 & 19.29 & 23.39 &  0.45 &  0.21 &  0.39 &   \nl 
V2346\dotfill  &  11.0116 & 41.4046 & 21.87 & 19.86 & 23.38 &  0.46 &  0.39 &  0.25 &   \nl 
V8443\dotfill  &  11.0880 & 41.4442 & 21.95 & 20.38 &\nodata&  0.35 &  0.27 &\nodata&
\enddata
\label{table:misc}
\tablecomments{Variables V14370 D31C was also found in Paper I.}
\end{planotable}
\end{small}

\subsection{Comparison with other catalogs}

The area of M31C field has not been observed frequently before and the
only overlapping variable star catalog is given by Magnier et
al.~(1997, hereafter Ma97). Out of 14 variable stars in Ma97 which are
located in our M31C field, we cross-identified 13. Of these 13 stars,
four (Ma97 79, 84, 88, 91) we did not classify as variables
($J_S=0.72, -0.04, 0.34, 0.43$). Of the remaining nine stars we have
classified eight as Cepheids and one as an eclipsing binary (see
Tables~\ref{table:ecl},\ref{table:ceph} for cross-ids).

There was also by design a slight overlap between the M31C and M31B
fields (Figure~\ref{fig:xy}). There were four Cepheids from the M31C
field in the overlap region, and they were all cross-identified in the
M31B catalog, with very similar properties of their light curves (see
Table~\ref{table:ceph}). There was only one eclipsing binary in the
overlap from the M31C field, V12594 D31C, and it was cross-identified
as V2763 D31B, again with very similar properties of its light curve
(see Table~\ref{table:ecl}).  We also cross-identified one
miscellaneous variable (see Table~\ref{table:misc}), out of three
detected in the M31B field and one detected in the M31C field, which
fell into the overlap region.

\begin{figure}[p]
\plotfiddle{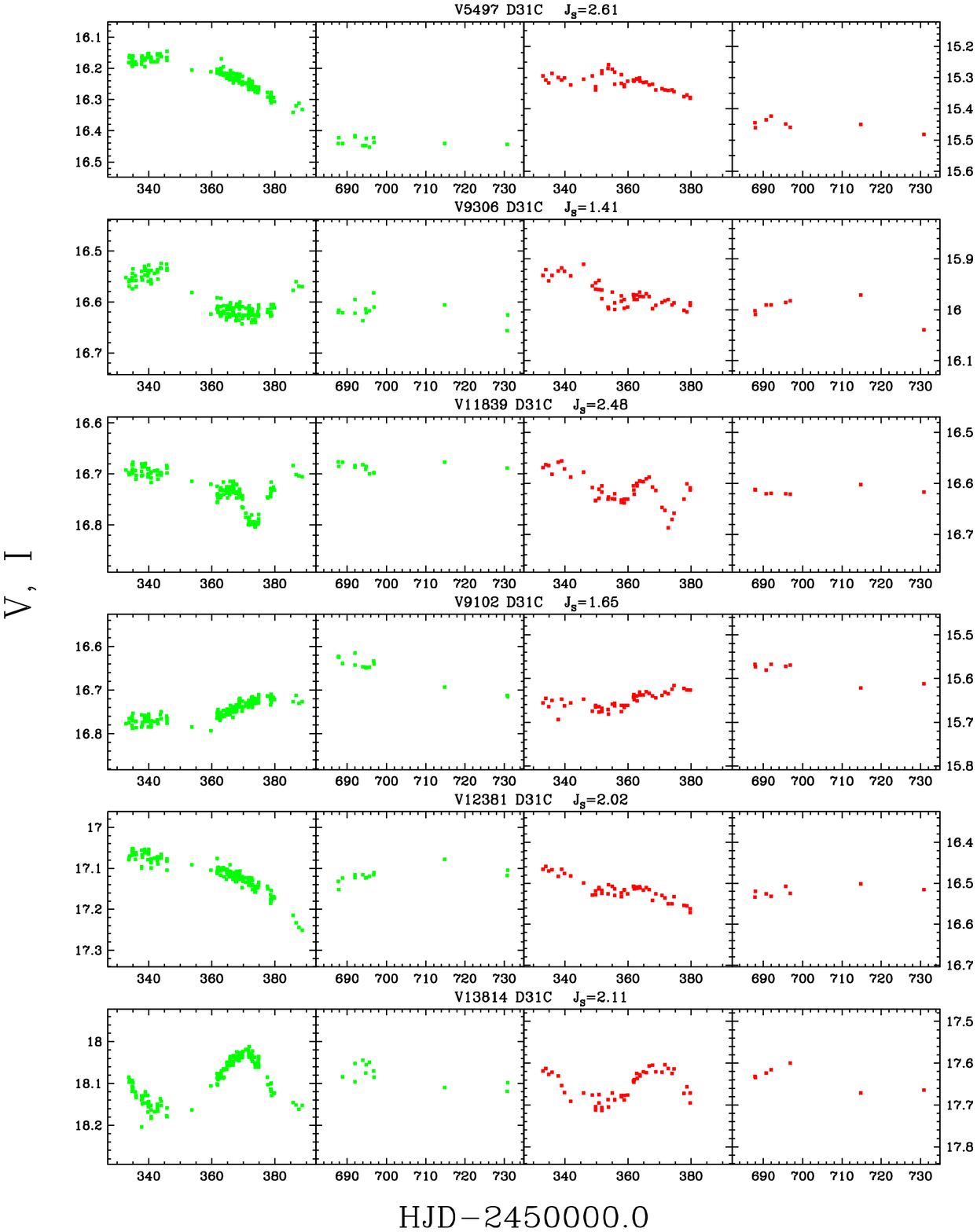}{19.5cm}{0}{83}{83}{-260}{-40}
\caption{$VI$ lightcurves of the miscellaneous variables found in the
field M31C.  $I$ (if present) is plotted in the two right panels.
$B$-band data is not shown.}
\label{fig:misc}
\end{figure}

\addtocounter{figure}{-1}
\begin{figure}[p]
\plotfiddle{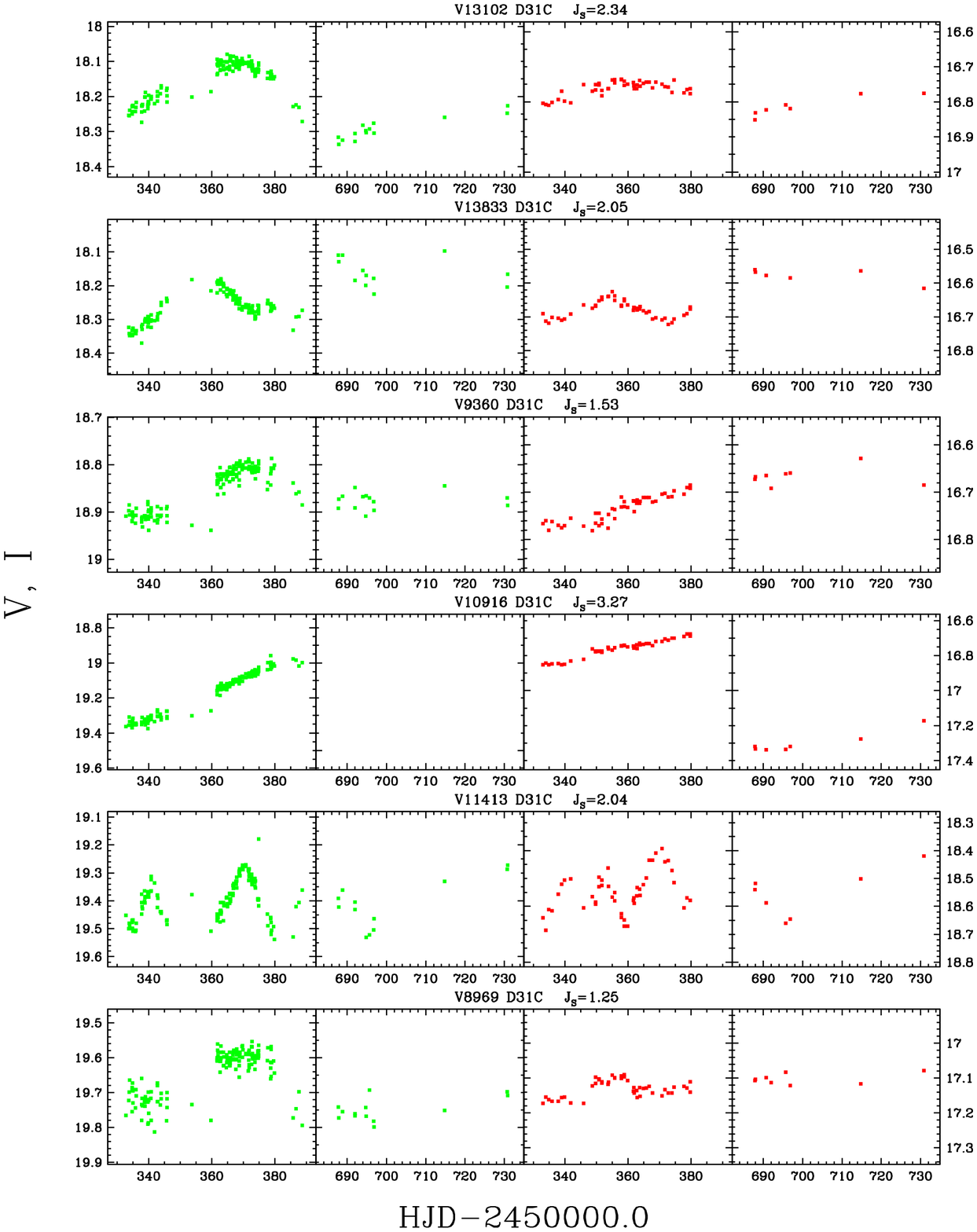}{19.5cm}{0}{83}{83}{-260}{-40}
\caption{Continued.}
\end{figure}

\addtocounter{figure}{-1}
\begin{figure}[p]
\plotfiddle{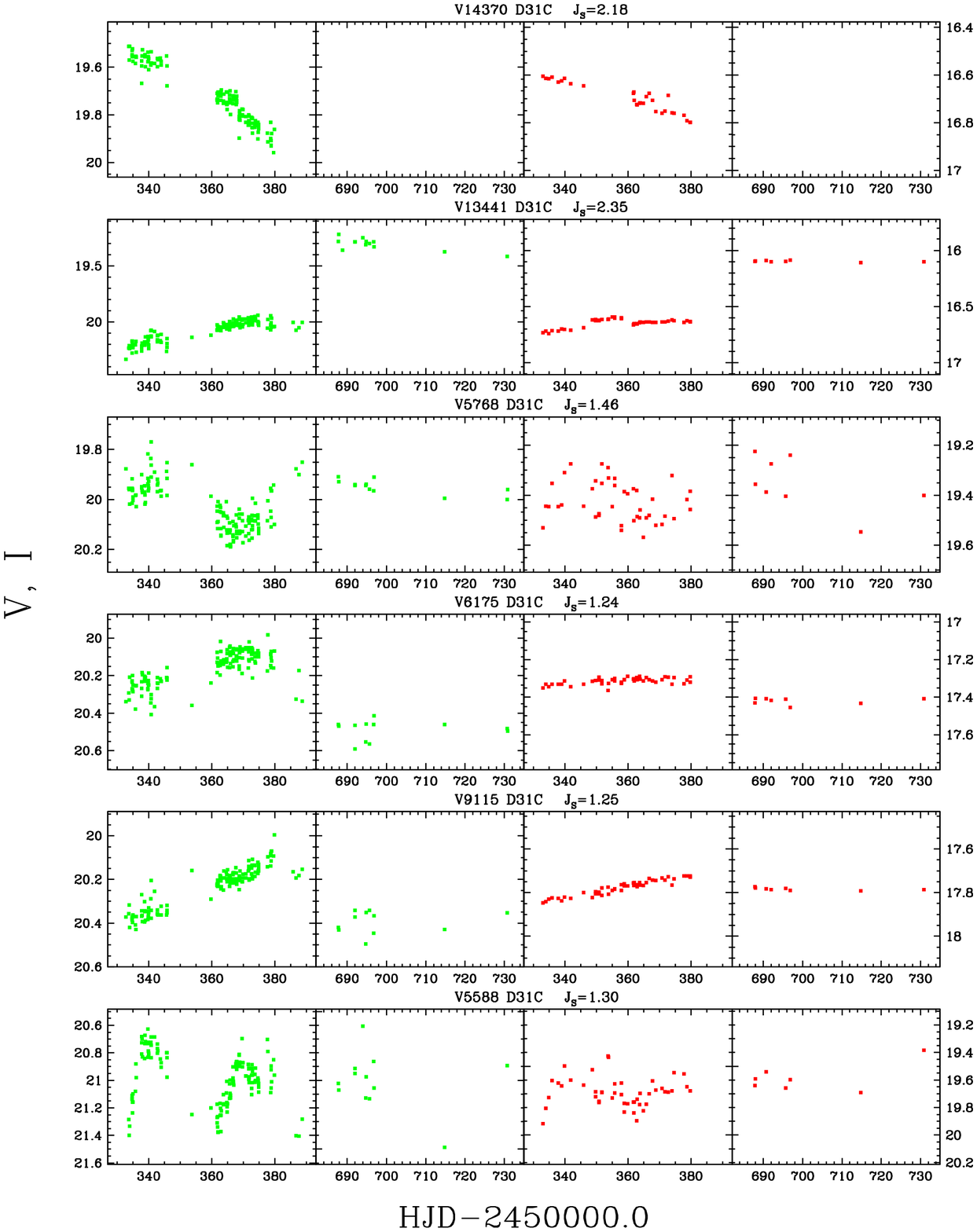}{19.5cm}{0}{83}{83}{-260}{-40}
\caption{Continued.}
\end{figure}

\addtocounter{figure}{-1}
\begin{figure}[p]
\plotfiddle{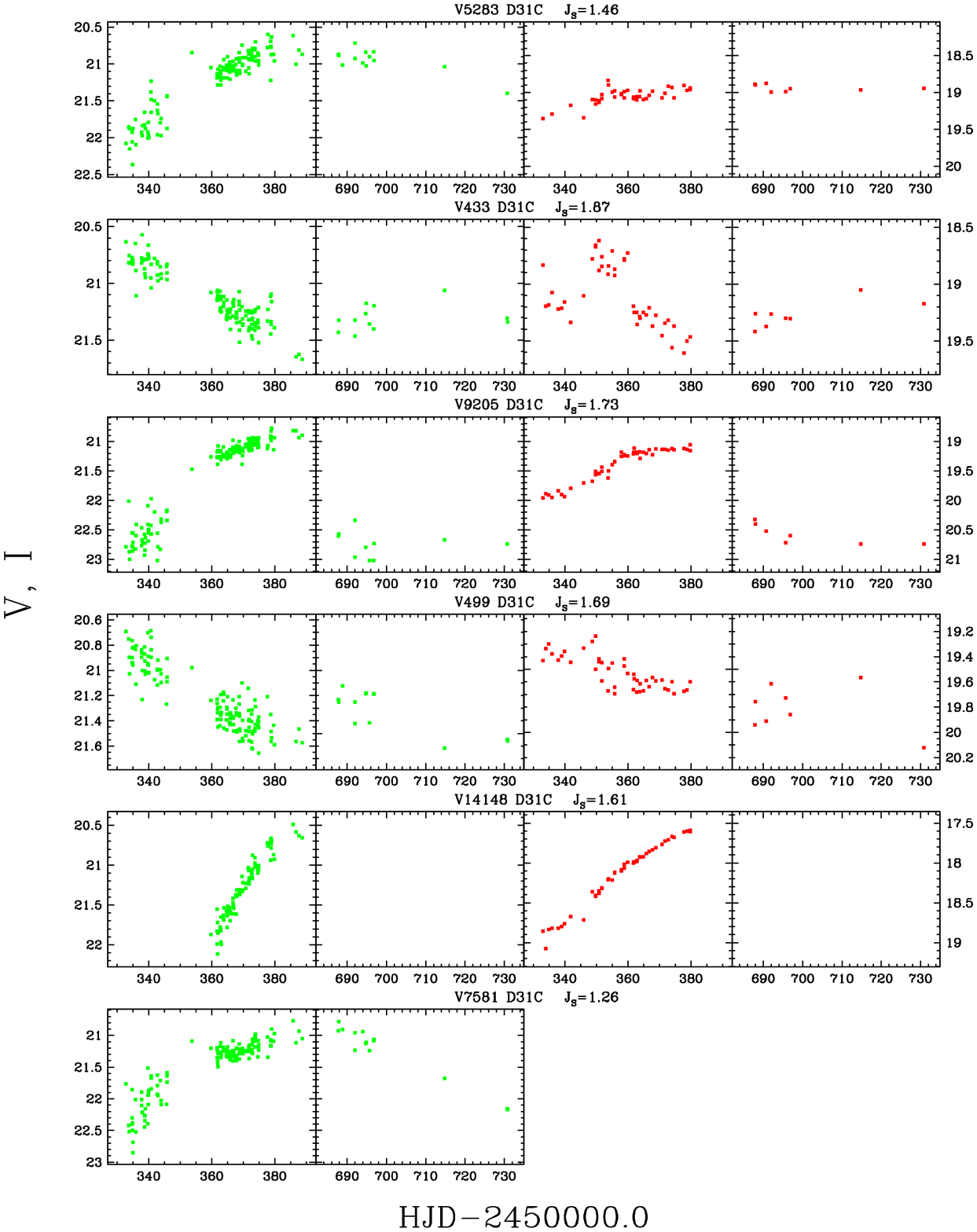}{19.5cm}{0}{83}{83}{-260}{-40}
\caption{Continued.}
\end{figure}

\addtocounter{figure}{-1}
\begin{figure}[p]
\plotfiddle{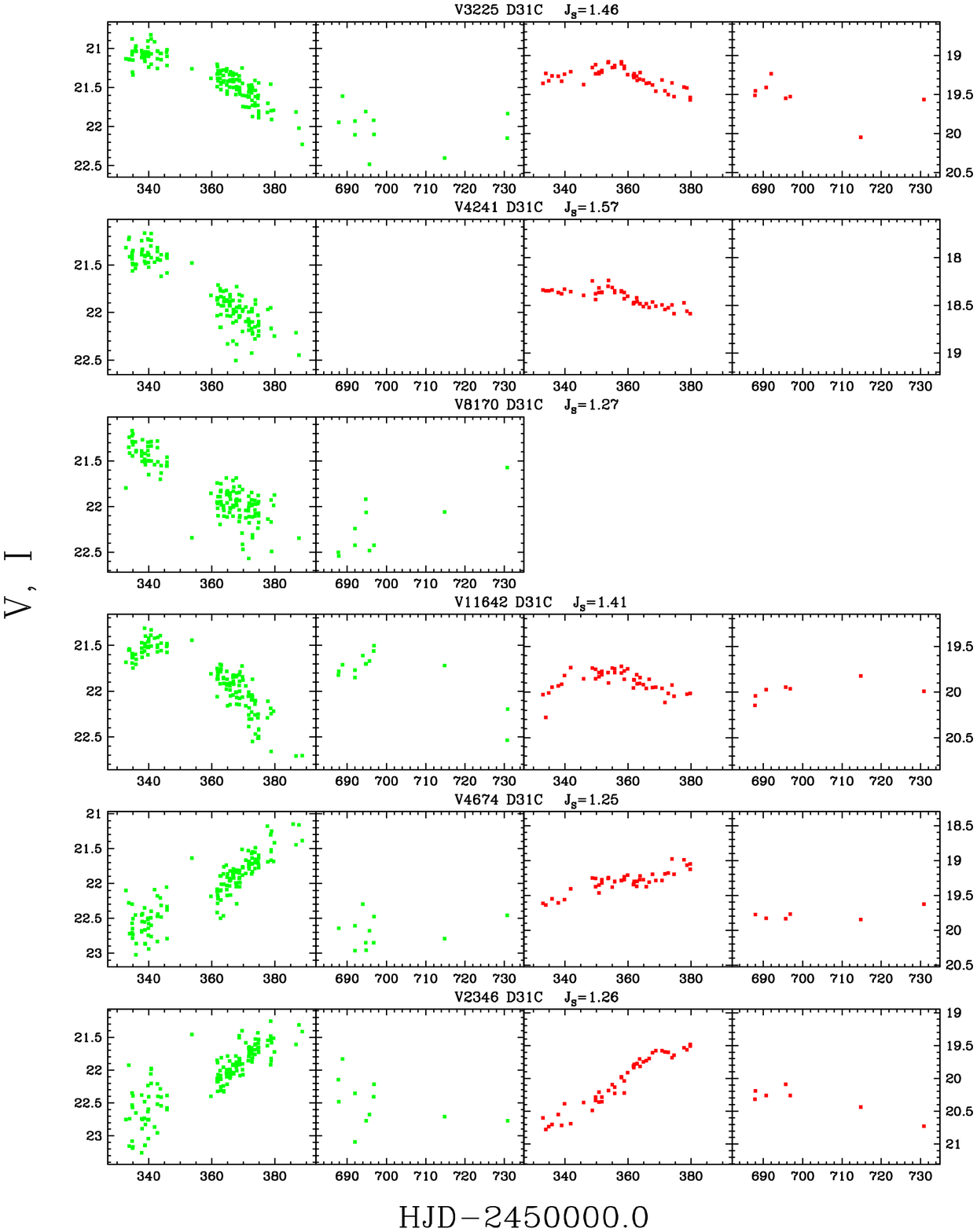}{19.5cm}{0}{83}{83}{-260}{-40}
\caption{Continued.}
\end{figure}

\addtocounter{figure}{-1}
\begin{figure}[p]
\plotfiddle{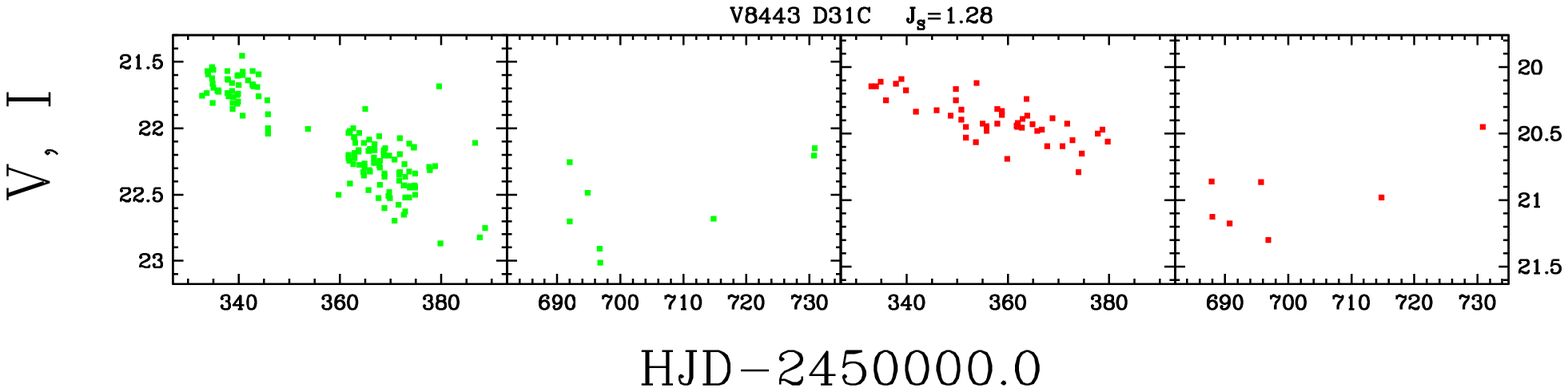}{5.5cm}{0}{83}{83}{-260}{-500}
\caption{Continued.}
\end{figure}

\section{Discussion}

\begin{figure}[t]
\plotfiddle{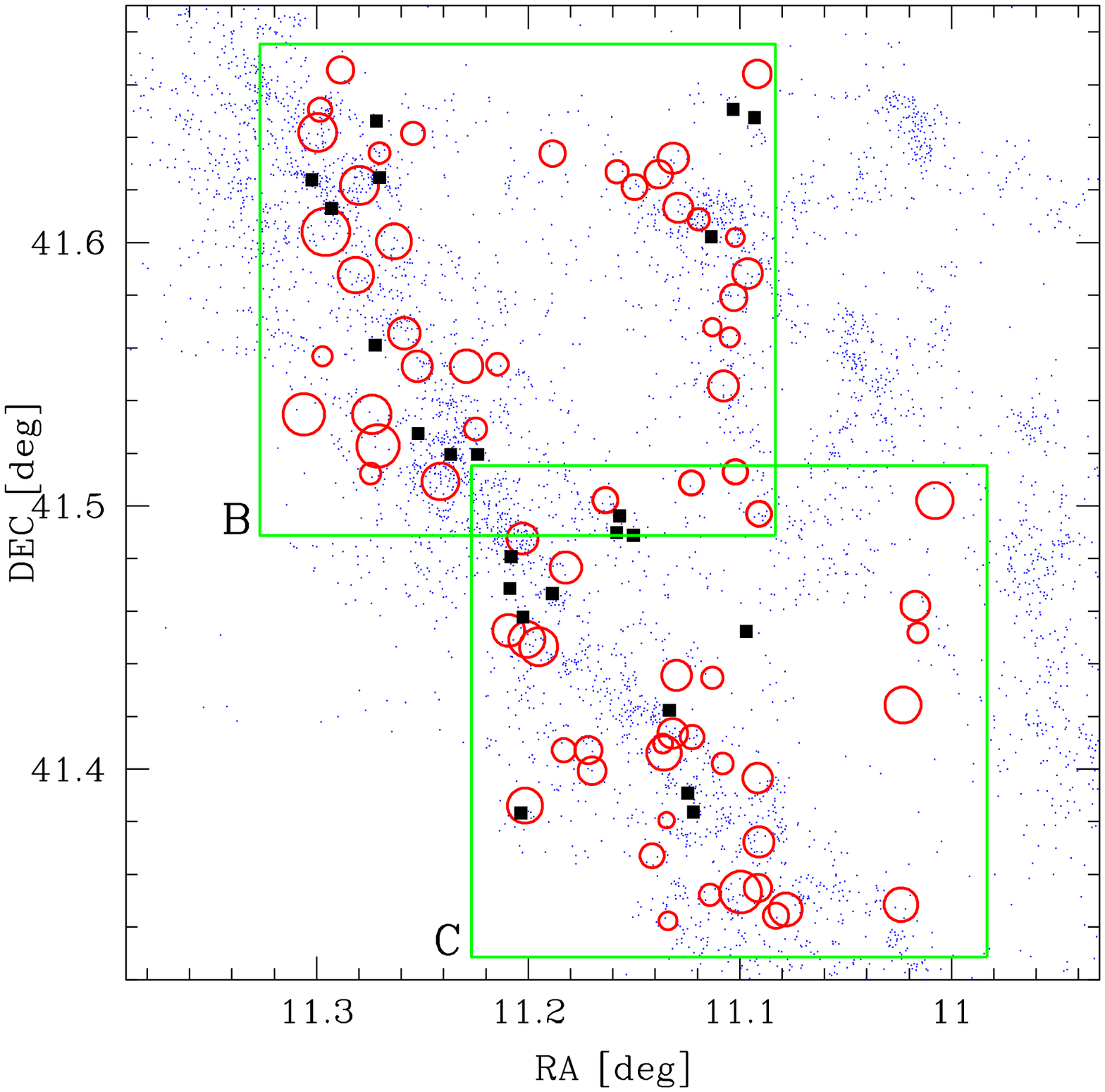}{8cm}{0}{50}{50}{-160}{-85}
\caption{Location of eclipsing binaries (filled squares) and Cepheids
(open circles) in the fields M31C and M31B, along with the blue stars
($B-V<0.4$) selected from the photometric survey of M31 by Magnier et
al.~(1992) and Haiman et al.~(1994). The sizes of the circles
representing the Cepheids variables are proportional to the logarithm
of their period.
\label{fig:xy}}
\end{figure}
 
In Figure~\ref{fig:cmd} we show $V,\;V-I$ and $V,\;B-V$
color-magnitude diagrams for the variable stars found in the field
M31C. The eclipsing binaries and Cepheids are plotted in the left
panels and the other periodic variables and miscellaneous variables
are plotted in the right panels.  As expected, most of the eclipsing
binaries occupy the blue upper main sequence of M31 stars, with the
exception of the bright, probably foreground, W UMa system V13944
D31C. The Cepheid variables group near $B-V\sim1.0$, with considerable
scatter probably due to differential reddening across the field. The
other periodic variable stars have positions on the CMD similar to the
Cepheids. The miscellaneous variables are scattered throughout the
CMDs and represent several classes of variability. Many of them are
very red with $V-I>2.0$, and are probably Mira variables. Several
brightest miscellaneous variables are probably foreground stars
belonging to our Galaxy.

In Figure~\ref{fig:xy} we plot the location of eclipsing binaries and
Cepheids in the fields M31C and M31B, along with the blue stars
($B-V<0.4$) selected from the photometric survey of M31 by Magnier et
al.~(1992) and Haiman et al.~(1994). The sizes of the circles
representing the Cepheids variables are proportional to the logarithm
of their period. As could have been expected, both types of variables
group along the spiral arms, as they represent relatively young
populations of stars.  We will explore various properties of our
sample of Cepheids in the future paper (Sasselov et al.~1999, in
preparation).

\acknowledgments{We would like to thank the TAC of the
Michigan-Dartmouth-MIT (MDM) Observatory and the TAC of the
F.~L.~Whipple Observatory (FLWO) for the generous amounts of telescope
time devoted to this project. We are very grateful to Bohdan
Paczy\'nski for motivating us to undertake this project and his always
helpful comments and suggestions.  We thank Lucas Macri for taking
some of the data described in this paper and Przemek Wo\'zniak for his
FITS-manipulation programs. The staff of the MDM and the FLWO
observatories is thanked for their support during the long observing
runs.  KZS was supported by the Harvard-Smithsonian Center for
Astrophysics Fellowship. JK was supported by NSF grant AST-9528096 to
Bohdan Paczy\'nski and by the Polish KBN grant 2P03D011.12. JLT was
supported by the NSF grant AST-9401519.}


\begin{references}

\reference{} Andersen, J. 1991, A\&AR, 3, 91
\reference{} Baade, W., Swope, H. H. 1963, AJ, 68, 435
\reference{} Baade, W., Swope, H. H. 1965, AJ, 70, 212 
\reference{} Crotts, A. P. S.,  \& Tomaney, A. B. 1996, ApJ, 473, L87
\reference{} Freedman, W. L., \& Madore, B. F. 1990, ApJ, 365, 186 
\reference{} Freedman, W. L., Wilson, C. D., \& Madore, B. F. 1991, 
	ApJ, 372, 455 
\reference{} Gaposchkin, S. 1962, AJ, 67, 334
\reference{} Haiman, Z., et al. 1994, A\&A, 286, 725 
\reference{} Hilditch, R. W. 1996, in: ASP Conf. Ser. 90, The Origins, Evolution
	and Destinies of Binary Stars in Clusters, ed. E. F. Milone \& 
	J.-C. Mermilliod (San Francisco: ASP), 207 
\reference{} Holland, S. 1998, AJ, 115, 1916
\reference{} Hubble, E. 1926, ApJ, 63, 236
\reference{} Hubble, E. 1929, ApJ, 69, 103 
\reference{} Huterer, D., Sasselov, D. D., Schechter, P. L. 1995, AJ, 100, 2705
\reference{} Jacoby, G. H., et al. 1992, PASP, 104, 599
\reference{} Kaluzny, J., Stanek, K. Z., Krockenberger, M., Sasselov, D. D.,
	Tonry, J. L., \& Mateo, M. 1998, AJ, 115, 1016 (Paper I)
\reference{} Krockenberger, M., Sasselov, D. D., \& Noyes, R. 1997, ApJ, 479, 875 
\reference{} Lafler, J., \& Kinman, T. D. 1965, ApJS, 11, 216
\reference{} Landolt, A. 1992, AJ, 104, 340 
\reference{} Magnier, E. A., Augusteijn, T., Prins, S., van Paradijs, J., \& 
        Lewin, W. H. G. 1997, A\&AS, 126, 401 (Ma97)
\reference{} Magnier, E. A., Lewin, W. H. G., Van Paradijs, J., Hasinger, G., 
        Jain, A., Pietsch, W., \& Truemper, J. 1992, A\&AS, 96, 37 
\reference{} Metzger, M. R., Tonry, J. L., \& Luppino, G. A. 1993, in 
	ASP Conf. Ser. 52, Astronomical Data Analysis Software and Systems II,
	ed. R. J. Hanisch, R. J. V. Brissenden, \& J. Barnes,
	(San Francisco: ASP), 300 
\reference{} Paczy\'nski, B. 1997, in The Extragalactic Distance Scale, 
	ed. M. Livio, M. Donahue \& N. Panagia
	(Cambridge: Cambridge Univ.~Press), 273
\reference{} Sasselov, D. D., et al. 1999, in preparation 
\reference{} Stanek, K. Z., Kaluzny, J., Krockenberger, M., Sasselov, D. D.,
	Tonry, J. L., \& Mateo, M. 1998, AJ, 115, 1894 (Paper II)
\reference{} Stanek. K. Z., \& Garnavich, P. M. 1998, ApJ, 503, L131
\reference{} Stetson, P.B. 1987, PASP, 99 191
\reference{} Stetson, P.B. 1992, in ASP Conf. Ser. 25, Astrophysical Data 
	Analysis Software and Systems I, ed. D. M. Worrall, 
	C. Bimesderfer, \& J. Barnes (San Francisco: ASP), 297 
\reference{} Stetson, P. B. 1996, PASP, 108, 851
\reference{} Szentgyorgyi, A., et al. 1999, in preparation 
\reference{} Tonry, J. L., Blakeslee, J. P., Ajhar, E. A., \& Dressler, A., 
        1997, ApJ, 475, 399 

\end{references}
\end{document}